\documentclass[twocolumn,aps,prd,amsmath,amssymb,preprintnumbers,longbibliography]{revtex4-1}
\usepackage{amsmath} \usepackage{amsfonts} \usepackage{amssymb}
\usepackage{bbm}
\usepackage{epsfig}
\usepackage{graphics}
\usepackage{graphicx}
\usepackage{titlesec}
\usepackage{mathtools}
\usepackage{environ}
\usepackage{dsfont}
\usepackage{version}
\usepackage[colorlinks=true]{hyperref}
\hypersetup{urlcolor=black,linkcolor=black,citecolor=black}
\usepackage{natbib}
\usepackage[toc,page]{appendix}
\usepackage{textalpha} 

\usepackage{enumerate}

\usepackage{dsfont}

\makeatletter
\renewcommand*\env@matrix[1][c]{\hskip -\arraycolsep
  \let\@ifnextchar\new@ifnextchar
  \array{*\c@MaxMatrixCols #1}}
\makeatother

\makeatletter										
\def\p@subsection{\thesection .\,} 
\makeatother                                       

\newcommand{\be}{\begin{equation}}
\newcommand{\ee}{\end{equation}}
\newcommand{\ba}{\begin{align}}
\newcommand{\ea}{\end{align}}
\newcommand{\nn}{\nonumber}

\newcommand{\gl}{\big(}
\newcommand{\gr}{\big)}

\newcommand{\vp}{{\varphi}}

\titleformat{\subsection}[block]{\normalfont\bfseries}{\thesubsection.}{1ex}{}
\titlespacing{\subsection}{0pt}{10pt}{1pt}[0pt]
\titleformat*{\section}{\large\bfseries}
\renewcommand{\thesubsection}{\arabic{subsection}}

\pdfoutput=1 

\usepackage{todonotes}
\usepackage{braket}

\newcommand{\bel}[1]{\be\label{#1}}

\newcommand{\Ktil}{\widetilde{K}}

\newcommand{\subt}[1]{_{\text{#1}}}
\newcommand{\supt}[1]{^{\text{#1}}}

\newcommand{\qq}[1]{``#1"}

\newcommand{\dt}{\partial_{t}}
\newcommand{\eps}{\varepsilon}

\newcommand{\gtil}{\tilde{g}}

\newcommand{\mtil}{\widetilde{m}}

\newcommand{\mo}{^{\mu}}

\newcommand{\mb}{_{\mu}}
\newcommand{\nb}{_{\nu}}
\newcommand{\mno}{^{\mu\nu}}
\newcommand{\mnb}{_{\mu\nu}}

\newcommand{\Mbar}{\overline{M}}
\newcommand{\rhotil}{\tilde{\rho}}

\newcommand{\M}{M^2}
\newcommand{\cH}{\mathcal{H}}
\newcommand{\cHhat}{\widehat{\cH}}
\newcommand{\deta}{\partial_\eta}
\newcommand{\Vhat}{\widehat V}
\newcommand{\Khat}{\widehat K}
\newcommand{\vptil}{\tilde\vp}
\newcommand{\cA}{\mathcal A}
\newcommand{\Pit}{\tilde\pi}
\newcommand{\Nbar}{\overline N}

\newcommand{\cN}{\mathcal{N}}
\newcommand{\Ttil}{\tilde{T}}
\newcommand{\der}[2]{\frac{\partial #1}{\partial #2}}
\newcommand{\sder}[2]{\frac{\partial^2 #1}{\partial #2^2}}
\newcommand{\Zbar}{\bar{Z}}
\newcommand{\ftil}{\tilde{f}}

\newcommand{\zwisch}[1]{\subsubsection*{\raggedright{\normalsize{\textit{#1}}}}}

\definecolor{refkey}{rgb}{0,0,1}
\definecolor{labelkey}{rgb}{0,1,0}

\begin{document}

\title{\LARGE Quantum gravity and scale symmetry in cosmology}

\author{C. Wetterich}

\affiliation{Institut  f\"ur Theoretische Physik\\
Universit\"at Heidelberg\\
Philosophenweg 16, D-69120 Heidelberg}

\begin{abstract}

We discuss predictions for cosmology which result from the scaling solution of functional flow equations for a quantum field theory of gravity. A scaling solution is necessary to render quantum gravity renormalizable. Our scaling solution is directly connected to the quantum effective action for the metric coupled to a scalar field. This effective action includes all effects of quantum fluctuations and is invariant under general coordinate transformations. Solving the cosmological field equations derived by variation of the quantum effective action provides for a detailed quantitative description of the evolution of the universe. The \qq{beginning state} of the universe is found close to an ultraviolet fixed point of the flow equation. It can be described by an inflationary epoch, with approximate scale invariance of the observed primordial fluctuation spectrum explained by approximate quantum scale symmetry. Overall cosmology realizes a dynamical crossover from the ultraviolet fixed point to an infrared fixed point which is approached in the infinite future. Present cosmology is close to the infrared fixed point. It features dynamical dark energy mediated by a light scalar field. The tiny mass of this cosmon arises from its role as a pseudo Goldstone boson of spontaneously broken quantum scale symmetry. The extremely small value of the present dark energy density in Planck units results dynamically as a consequence of the huge age of the universe. The cosmological constant problem finds a dynamical solution. We present a detailed quantitative computation of the scaling solution for the scalar effective potential and the field-dependent coefficient of the curvature scalar. This allows for further quantitative predictions.

\end{abstract}

\maketitle

\section{Introduction}\label{sec: I}

Cosmology is a testing ground for quantum gravity. First of all this concerns the beginning of the universe which is often associated in classical gravity to a \qq{big bang singularity}. In this very early epoch quantum gravity effects induced by fluctuations of the metric are generally thought to be important. Very early cosmology is often described by an inflationary epoch~\cite{STA, GUT, MUK, ALI, SWI} whose properties depend crucially on the shape of the potential for a scalar field. One may wonder if quantum gravity permits to compute this \qq{inflaton potential} or, at least, some of its important qualitative properties. The shape of this potential can be tested by indirect observations of the primordial fluctuation spectrum through the observed anisotropies of the cosmic microwave background (CMB). Quantum gravity predictions for the inflaton potential can be falsified by CMB observations, constituting important tests for a given approach or model. We will see that quantum gravity may even predict the shape of the scalar potential in a field region that is relevant for present dynamical dark energy or quintessence~\cite{Wetterich_1988, RP1, RP2, CWCMAV, FTSW, FEJO, VL, CLW, CDS, LA1, LACQ, LIN}. In this case the detailed observations of the properties of dark energy constitute further tests of a quantum gravity model.

In this work we focus on the formulation of quantum gravity as a quantum field theory for the metric. Further degrees of freedom are the fields for the standard model of particle physics and beyond, and a scalar singlet field that can play the role of the inflaton in early cosmology or the cosmon for late cosmology. A quantum field theory of gravity is \qq{renormalizable} and \qq{ultraviolet complete} if it admits an ultraviolet (UV) fixed point. Such a fixed point requires the existence of a \qq{scaling solution} for the functional flow equations describing the scale-dependence of couplings or coupling functions. At a fixed point the physics becomes independent of any length scale and can therefore be extrapolated to arbitrarily short distances. If interactions are present at the UV-fixed point a theory is called \qq{asymptotically safe}~\cite{WEIN}, otherwise it is \qq{asymptotically free}~\cite{GWI, POL}.

The main reason for our focus is that modern functional renormalization group techniques~\cite{Wetterich_1993, RWEAEE, MR}, see ref.~\cite{BEGREV} for a recent review, permit a detailed quantitative computation of the flow equations for quantum gravity, and therefore a detailed understanding of the scaling solutions. Quantum fluctuations of the metric are found to play an important role for many properties of the scalar effective potential or similar coupling functions. The effect of the metric fluctuations is described quantitatively and permits important predictions for cosmology. So far other approaches to quantum gravity do not yet yield a sufficient quantitative understanding of the effects of quantum fluctuations that would allow for a meaningful comparison with our quantitative results. Either fluctuation effects of the metric are very difficult to be incorporated, especially in the non-perturbative region relevant for fixed points. This is the case for string theories. Or it is difficult to formulate diffeomorphism invariant field equations for some type of metric field or a similar \qq{geometric field}. Such field equations are crucial for a quantitative description of cosmology. This present shortcoming is typically given for lattice approaches to quantum gravity. For some other approaches the issue may simply be the lack of present computational capability. Waiting for further developments of alternative approaches we are aware that the limitations of the present work do not do justice to many interesting qualitative arguments and conjectures of these approaches for the beginning stage of the universe. We also remain strictly within the setting of a diffeomorphism invariant effective action and the field equations following from it. This restriction omits many interesting proposals for \qq{shortcuts} by identifying the renormalization scale $k$ of functional flow equations with some geometrical quantity, see the review~\cite{PLA1} and references therein.

A central ingredient for the present work is the scaling solution for the scalar effective potential and the field- and scale-dependent \qq{curvature coefficient} or \qq{effective Planck mass}~\cite{HPW, HPRW, CWQS, PRWY, CWMY, CWESPA}. Within our approximations we find that such a scaling solution exists and permits gravity to be a renormalizable quantum field theory. We compute the scaling solutions quantitatively. The most predictive scenario is realized by \qq{fundamental scale invariance}~\cite{CWFSI} for which the scaling solution directly describes the quantum effective action. For a more general renormalizable quantum field theory a small number of \qq{relevant parameters} describes the flow away from the UV-fixed point as the renormalization scale is lowered towards the infrared. The presence of these additional free parameters reduces somewhat the predictive power for some of the quantitative results. The overall picture of cosmology remains similar, however.

Our approach leads to several key predictions for cosmology:
\begin{enumerate}[(i)]
\item The beginning epoch of the universe can be described by inflationary cosmology.
\item The cosmological constant problem finds a dynamical solution.
\item The overall history of the universe is a crossover from the vicinity of a UV-fixed point in the infinite past to an IR-fixed point in the infinite future. The approximate quantum scale symmetry near the UV-fixed point explains the approximate scale invariance of the primordial cosmic fluctuation spectrum.
\item The approach to the IR-fixed point realizes some form of dynamical dark energy.
\item The quantum scale symmetry at the IR-fixed point is broken spontaneously, inducing a massless Goldstone boson. Close to the fixed point a very small mass for the cosmon - the pseudo Goldstone boson of spontaneously broken approximate quantum scale symmetry - is induced by explicit symmetry breaking of dilatation symmetry.
\item The tiny ratio $U/M^4\approx 10^{-120}$ of the present dark energy density $\sim U$ over the fourth power of the Planck mass $M$ can find a simple explanation by the huge age of the universe in Planck units. This is similar to a similar tiny ratio for the matter energy density or radiation energy density.
\end{enumerate}

Further more detailed quantitative predictions for fundamental scale invariance will be developed in the main text and are summarized in the conclusions.

We concentrate in this work on an approximation of the exact flow equation~\cite{Wetterich_1993, RWEAEE, CWTET, ELL, MOR} which consists in truncating the flowing effective action or effective average action for the metric and scalar field to the most general diffeomorphism invariant form containing up to two derivatives of the fields. This truncation involves three functions of the scalar field: the effective potential, the \qq{curvature coefficient} which multiplies the term linear in the curvature scalar $R$, and the \qq{kinetial} or \qq{wave function renormalization} which multiplies the kinetic term of the scalar field. Within this truncation we present explicit computations for the effective potential and the curvature coefficient, while similarly robust results for the kinetial are not yet available. The truncation to two derivatives may be expected to be valid if typical momenta are sufficiently small as compared to the effective Planck mass. This seems to be realized for the late stages of inflation relevant for the observable primordial fluctuation spectrum and for all later epochs of the universe. Towards the beginning in the early stages of inflation terms with more than two derivatives may become more important. This holds, in particular, if quantum gravity is asymptotically free~\cite{STE, FRATSE, AB, SWY} with dominant terms quadratic in the curvature tensor involving four derivatives. It is also possible that the beginning stage is better described in terms of other degrees of freedom, for example by gauge fields and a vierbein in \qq{pregeometry}~\cite{CWPC}.

In quantum field theory the observables do not depend on the particular choice of fields used to describe them. In particular, the choice of the metric is not unique. It may be changed by multiplication with a function of the scalar field, the so called \qq{Weyl scaling}~\cite{HWGE, RDMPI}. Different choices of the metric correspond to different \qq{metric frames}. The physics expressed in terms of observables is independent of the choice of the metric frame. We present many results directly in terms of frame invariant equations~\cite{CWCFVG, JIQG}. Nevertheless, for making contact with intuition and facilitating comparison with the existing literature it is useful to discuss a Weyl transformation to the \qq{Einstein frame} for which the Planck mass takes a fixed value. This fixed Planck mass is not an intrinsic scale of the quantum gravity model, being introduced only by a field transformation. One should not be surprised that many simple findings associated to fixed points and quantum scale symmetry get obscured if an \qq{artificial mass scale} is introduced. This explains why the naturalness of some of our results is not easily seen in the Einstein frame, for which too simple estimates would judge them as unnatural. This concerns, in particular, the properties of the IR-fixed point which lead to an effective potential that vanishes naturally for large field values and the associated very small mass of the scalar field responsible for dynamical dark energy.

This work is organized as follows: In sect.~\ref{sec:CBEG} we briefly discuss the need for cosmology beyond Einstein gravity which motivates to consider a scalar field along with the metric. Sect.~\ref{sec:QG} turns to the formulation of quantum gravity as a quantum field theory for the metric and a scalar field. It explains basic notions as the flow equations and the scaling solution. It also presents first qualitative results on a simple level. In sec.~\ref{sec:QSS} we emphasize the crucial role of quantum scale symmetry for the flow close to the UV- and IR-fixed points. This provides already for an overall picture of cosmology as a crossover from the UV- to the IR-fixed point, connecting an early inflationary epoch to late cosmology with dynamical dark energy. Sect.~\ref{sec:FESSQG} is devoted to quantitative results for the flow equations and we discuss the corresponding scaling solution in sect.~\ref{sec:SS2}. On this basis we discuss the crossover cosmology associated to the scaling solution in more quantitative detail in sect.~\ref{sec:CC}. We describe how the sequence of different epochs in the evolution of the universe emerges naturally for our setting of quantum gravity. In sect.~\ref{sec:FSI} we focus on fundamental scale invariance, shedding more light on which predictions arise only for this particular setting while more freedom is left for a general renormalizable quantum field theory of gravity. In sect.~\ref{sec:D} we summarize our results.

\section{Cosmology beyond Einstein gravity}\label{sec:CBEG}

General relativity or Einstein gravity is a classical field theory, based on the Einstein-Hilbert action
\bel{C1}
S=\int_x\sqrt{g}\bigg\{-\frac{\M}{2}R+V\bigg\}\ .
\ee
Here $R$ is the curvature scalar formed from the metric $g\mnb(x)$ and $g=\text{det}\gl g\mnb\gr$. (The factor $\pm i$ for $g<0$ plays no role for the field equations.) The integral $\int_x$ is an integral over four-dimensional space. This theory involves two parameters, the (reduced) Planck mass $M=2.436\cdot10^{18}\,\text{GeV}$ and the cosmological constant $V=(2\cdot10^{-3}\,\text{eV})^4$. Gravity is characterized by a fundamental symmetry, namely diffeomorphism symmetry, which is equivalent to invariance under general coordinate transformations. Infinitesimal diffeomorphism transformations can be formulated as variations of the metric at fixed coordinates, with infinitesimal parameters $\xi\mo(x)$,
\bel{C2}
\delta g\mnb=-\partial\mb\xi^\rho g_{\rho\nu}-\partial\nb\xi^\rho g_{\mu\rho}-\xi^\rho\partial_\rho g\mnb\ .
\ee
The gravitational field equations obtain by variation of $S$ with respect to $g\mnb(x)$.

Together with a minimal coupling of the metric to matter fields, both through covariant derivatives and the overall factor $\sqrt{g}$, Einstein gravity is an extremely successful model for most observations in gravity and cosmology. It describes the gravity of stars and black holes, as well as galaxies or larger structures once some form of dark matter is included. It is tested with high precision by experiments from the submillimeter scale to the size of our solar system. For a suitable matter content of the universe Einstein gravity successfully describes the hot radiation dominated universe and the subsequent matter dominated universe, as tested by nucleosynthesis or the cosmic microwave background radiation (CMB). In presence of the cosmological constant $V$ it can account for dark energy and the associated accelerated expansion in the recent cosmological epoch.

Nevertheless, Einstein gravity has also a few important shortcomings.
\begin{enumerate}[(i)]
\item The classical theory has to be extended to quantum gravity in order to permit a consistent coupling to quantum matter.
\item The tiny dimensionless ratio $V/M^4\approx10^{-120}$ remains unexplained.
\item The initial value for the energy density must be extremely close to the critical density, which needs tremendous fine tuning.
\item The high degree of isotropy of the CMB remains a mystery, since within Einstein gravity the radiation from different directions (angles larger than $\sim1$°) are emitted from regions that never had causal contact.
\item The singularities in the center of the black hole or at the beginning of the universe may point to an incompleteness of this theory.
\item A two parameter model for dark energy is very predictive and may be falsified if the present observational tensions (Hubble tension etc.) evolve to hard contradictions.
\end{enumerate}

Many (if not all) of the shortcomings may be cured by a quantum field theory for the metric coupled to a scalar field $\chi$. This scalar field is a singlet with respect to the gauge group of the standard model. It can play the role of the inflaton for an early inflationary epoch of the universe~\cite{STA, GUT, MUK, ALI, SWI}. Inflation can explain the closeness to the critical energy density and many properties of the primordial fluctuations which are observed through the anisotropies of the CMB. For the present cosmology the scalar field can be associated to the cosmon, a very light scalar field responsible for dynamical dark energy~\cite{Wetterich_1988, RP1, RP2, CWCMAV, FTSW, FEJO, VL, CLW, CDS, LA1, LIN, LACQ} and associated to a dynamical explanation for the tiny ratio between the dark energy density and $M^4$ around $10^{-120}$. It is possible that the same scalar field accounts for inflation and dynamical dark energy~\cite{SPO, PEVI, PERO, DIVA, GIO, BRMA, CWCI, HMSS, HMSS2, WHMSS, GUE, RUCW, HASO, BR}. A quantum field theory includes quantum fluctuations of both the scalar and the metric field. Generalizations may replace the metric by the vierbein or introduce other \qq{pregeometric} fields~\cite{CWPC}. We remain for this note with the metric. We also do not discuss the interesting possibility of asymptotically free quantum gravity which involves higher order curvature invariants~\cite{STE, FRATSE, AB, SWY}.

The central quantity for discussing cosmology in a quantum field theory is the quantum effective action $\Gamma[g\mnb,\chi]$. Here $g\mnb(x)$ corresponds to the expectation value of the fluctuating metric field, and $\chi$ is the expectation value of a fluctuating scalar field. Formally, the functional $\Gamma[g\mnb,\chi]$ generates the one-particle irreducible Green's functions. The first variation of $\Gamma$ with respect to $g\mnb$ or $\chi$ yields the exact quantum field equations, possibly with a source term arising from the matter part that we do not discuss here explicitly. These field equations constitute the relevant evolution equations for cosmology. The second functional variation yields the inverse correlation function. The correlation functions for the scalar and the metric field encode the primordial scalar and tensor fluctuations~\cite{CWMF, CWCFVG}. In order to establish the direct connection between $\Gamma$ and the observable quantities a diffeomorphism invariant form of the quantum effective action is needed~\cite{CWPM}. The aim of quantum gravity is the computation of $\Gamma$ by including all effects of fluctuations of the metric and other fields.

Before reporting on progress for the computation of $\Gamma$ it is useful to discuss a few general properties. Diffeomorphism symmetry requires the invariance of $\Gamma$ with respect to the transformation~\eqref{C2}, combined with $\delta\chi=-\xi^\rho\partial_\rho\chi$. We also assume a discrete symmetry $\chi\to-\chi$. For low enough (covariant) momenta and small enough curvature invariants one expects the validity of a derivative expansion. Up to second order in the derivatives the effective action in the gravity-scalar sector takes the form
\bel{C3}
\Gamma=\int_x\sqrt{g}\bigg\{-\frac12F(\chi)R+\frac12K(\chi)\partial\mo\chi\partial\mb\chi+U(\chi)\bigg\}\ .
\ee
This has to be supplemented by the effective action for the fields of the standard model of particle physics. The effective scalar potential $U(\chi)$ generalizes the cosmological constant $V$. In the presence of other scalar fields as the Higgs scalar this potential is supposed to describe the relative minimum with respect to the other fields. The \qq{kinetial} $K(\chi)$ multiplies the scalar kinetic term, with $\partial\mo\chi=g\mno\partial\nb\chi$ and $g\mno g_{\nu\rho}=\delta\mo_{\ \rho}$. Finally, the curvature coefficient $F(\chi)$ generalizes $\M$. Its $\chi$-dependence yields a modification of general relativity which may be named \qq{variable gravity}~\cite{CWVG}.

Unless dimensionless coefficients of higher derivative invariants as $R^2$, $R\mno R\mnb$, $\gl\partial\mo\chi\partial\mb\chi\gr^2/F^4$ etc. are much larger than one, the effective action~\eqref{C3} provides for a good approximation to the epoch of inflation relevant for the observed fluctuations, as well as for all later cosmology. Furthermore, a rather large class of modified gravity theories with higher derivatives can be brought to the form~\eqref{C3} by using appropriate fields and variable transformations~\cite{CWMODG}. This includes Starobinsky inflation~\cite{STA} which involves a very large coefficient of the higher derivative invariant $R^2$.

The field equations obtained by variation of the effective action for variable gravity~\eqref{C3} can be found in ref.~\cite{CWVG}. We are mainly interested here in homogeneous and isotropic solutions which correspond to a scalar field $\chi(\eta)$ only depending on conformal time $\eta$, and a metric $g\mnb=a^2(\eta)\eta\mnb$, with $a(\eta)$ the cosmic scale factor and $\eta\mnb=\text{diag}\gl-1,1,1,1\gr$. It is convenient to introduce
\bel{C4}
A=\sqrt{F}a\ ,\quad \cHhat=\deta\ln A=\cH+\frac12\deta\ln F\ ,
\ee
with conformal Hubble parameter $\cH=\deta\ln a=Ha$ related to the Hubble parameter $H=\partial_t\ln a$, and cosmic time $t$ related to $\eta$ by $\text{d}t=a\text{d}\eta$. We can write~\cite{CWCFVG} the homogeneous gravitational field equations for spatially flat geometries in the simple form
\bel{C5}
2\cHhat^2+\deta\cHhat=A^2\Vhat\ ,\quad \cHhat^2-\deta\cHhat=\frac{\Khat}2\gl\deta\chi\gr^2\ ,
\ee
while the scalar field equation reads
\bel{C6}
\Khat\gl\deta^2+2\cHhat\deta\gr\chi+\frac12\frac{\partial\Khat}{\partial\chi}\gl\deta\chi\gr^2+A^2\frac{\partial\Vhat}{\partial\chi}=0\ .
\ee
As usual, only two of the three equations~\eqref{C5},~\eqref{C6} are independent.

These equations employ the combinations
\bel{C7}
\Vhat=\frac U{F^2}\ ,\quad \Khat=\frac KF+\frac3{2F^2}\bigg(\frac{\partial F}{\partial\chi}\bigg)^2\ .
\ee
Out of the three functions $F$, $K$, and $U$ only the two particular combinations $\Vhat$ and $\Khat$ matter. We will later see that eqs.~\eqref{C5}-~\eqref{C7} are valid in all metric frames related by conformal or Weyl scalings, with frame invariant combinations $\Vhat$ and $\Khat$. For the special case $F=\M$, $K=1$ they reduce to the well known cosmological equations
\begin{align}
\label{C8}
H^2=&\frac{1}{3\M}\Big[U+\frac12\gl\partial_t\chi\gr^2\Big]\ ,\nn\\
\partial_t^2\chi&+3H\partial_t\chi=-\frac{\partial U}{\partial\chi}\ .
\end{align}

\section{Quantum gravity}\label{sec:QG}

The usual approach to cosmology beyond Einstein gravity assumes a particular form of the functions $U(\chi)$, $F(\chi)$ and $K(\chi)$, solves the field equations, and discusses consequences for observations. It should be the aim of a theory of quantum gravity to compute these functions, or at least to restrict their qualitative behavior. The transition from assumption to computation can be a major step in our understanding of cosmology. Using functional flow equations~\cite{Wetterich_1993, RWEAEE, MR, bonanno2020critical} we will find that the fluctuations of the metric indeed imply important restrictions and new features for the shape of the functions $U$, $F$, $K$. These effects are typically non-perturbative. It would be very important to compare these qualitative features with other approaches to quantum gravity. This would require other methods that can cope with the effects of metric fluctuations, including the non-perturbative domain. Unfortunately, such alternative methods seem not yet to be available.

\subsection{Flow equation}\label{subsec:FE}

A complete model of quantum gravity should be valid for all distance or momentum scales. If it does not involve a fundamental smallest length, the model should describe what happens as length scales approach zero or momenta reach infinity. The physics at different momentum scales typically changes due to the effects of quantum fluctuations. Effective laws at a given momentum type scale $k$ may be encoded in a scale dependent effective action $\Gamma_k$. More precisely, we define $\Gamma_k$ by only including the quantum fluctuations with squared (covariant) momenta $|q^2|\gtrsim k^2$. In the limit $k\to0$ all fluctuations are included, such that for $k\to0$ the scale dependent effective action $\Gamma_k$ equals the quantum effective action. In the opposite limit $k\to\infty$ no fluctuations are included, and $\Gamma_k$ approaches the microscopic (\qq{classical}) action that is used to define the functional integral for a quantum field theory. If it is possible to find a valid form of the functional $\Gamma_k$ for the whole range of $k$ from infinity to zero, a quantum field theory can be considered to be complete. In this case the model defined at an arbitrary small microscopic length scale $k^{-1}\to0$ can be related to the observable \qq{macrophysics} for $k\to0$.

The $k$-dependence of $\Gamma_k$ is described by a functional flow equation
\bel{Q1}
k\partial_k\Gamma_k=\zeta_k\ ,
\ee
where the flow generator $\zeta_k$ is a functional of $g\mnb$ and $\chi$. It typically involves an integral over the momenta of the fluctuations. We will work here with euclidean momenta $q^2\geq0$, with analytic continuation to Minkowski signature done at the end. Since in the step from $k+\text{d}k$ to $k$ only a small range of additional fluctuations is included, this momentum integral is finite, centered around $q^2\approx k^2$. We will employ a particular form of the flow equation based on a gauge invariant formulation of the effective average action~\cite{CWPM, CWFSI}. The contribution of low momentum fluctuations is removed by a smooth infrared cutoff function $R_k(q^2)$ which vanishes rapidly for $q^2\gg k^2$. Besides its dependence on the infrared cutoff the flow generator $\zeta_k$ only involves $\Gamma_k^{(2)}[g\mnb,\chi]$, the second functional derivative of $\Gamma_k$ evaluated for arbitrary fields $g\mnb$ and $\chi$. Thus both sides of eq.~\eqref{Q1} are functionals of these fields. The functional flow equation for the effective average action is an exact identity~\cite{Wetterich_1993}. For practical purposes it has to be approximated by truncating the most general form of $\Gamma_k$, for example to the form~\eqref{C3}. We will present more details in sect.~\ref{sec:FESSQG}. Here we first address the most important features and results.

\subsection{Scaling solution}\label{subsec:SS1}

For a scaling solution $\Gamma_k$ becomes independent of $k$ once it is expressed in terms of suitable dimensionless renormalized fields and coupling functions. In our approximation this means that the dimensionless functions
\bel{Q2}
u(\rhotil)=\frac{U}{k^4}\ ,\quad f(\rhotil)=\frac{F}{k^2}\ ,\quad K(\rhotil)\ ,
\ee
depend only on the dimensionless combination
\bel{Q3}
\rhotil=\frac{\chi^2}{2k^2}\ .
\ee
For a scaling solution the effective average action expressed in terms of these functions solves the flow equation for the whole range of $\rhotil$ from zero to infinity. In other words, the (truncated) flow equation~\eqref{Q1} should admit a solution for which the only $k$-dependence arises implicitly through the expressions~\eqref{Q2},~\eqref{Q3}, without any additional explicit $k$-dependence.

The existence of a scaling solution for $\Gamma_k$ implies that quantum gravity can be formulated as a complete quantum field theory. Indeed, for any finite non-zero $\chi$ we can extrapolate $\Gamma_k$ arbitrarily far to the ultraviolet, $k\to\infty$, by taking the limit $\rhotil\to0$. The infrared limit $k\to0$ corresponds to $\rhotil\to\infty$.

If $u(\rhotil)$ is analytic at $\rhotil=0$ the scaling solution corresponds to fixed points for infinitely many dimensionless renormalized couplings. We may define those couplings by a Taylor expansion of $u(\rhotil)$ at $\rhotil=0$,
\begin{align}
\label{Q4}
u(\rhotil)=&u_0+\mtil_0^2\rhotil+\frac12\lambda_0\rhotil^2+\frac16\tilde\gamma_0\rhotil^3+\dots\ ,\nn\\
U(\chi)=&u_0k^4+\frac{\mtil_0}{2}k^2\chi^2+\frac{\lambda_0}{8}\chi^4+\frac{\tilde\gamma_0}{48k^2}\chi^6+\dots\ .
\end{align}
For a scaling solution the flow of $u_0$, $\mtil_0^2$, $\lambda_0$, $\tilde\gamma_0$, $\dots$ becomes independent of $k$. These considerations extend to other functions that characterize the scaling solution for the functional $\Gamma_k$. In simple words, nothing changes anymore if the ultraviolet limit $k\to\infty$ is formulated in terms of renormalized dimensionless couplings. The theory is then ultraviolet complete. The existence of an ultraviolet fixed point at non-zero couplings is the basic idea of asymptotic safety for quantum gravity~\cite{WEIN, MR, SOUM, DPER, RSAU, LAUR}.

We emphasize that the existence of a scaling solution does not require that the functions $u(\rhotil)$, $f(\rhotil)$, $K(\rhotil)$ remain all finite for $\rhotil\to0$. We have formulated the flow equation for one given field basis $\gl g\mnb,\chi\gr$. It is possible that in this field basis some functions, say $K(\rhotil)$, diverges for $\rhotil\to0$. By non-linear field transformations one may find a different choice of fields (\qq{different metric frame}) for which the dimensionless functions remain finite for $k\to\infty$. For the existence of a scaling solution it is sufficient that one choice of fields exists for which $\Gamma_k$ is well defined for all field values and shows no explicit $k$-dependence.

\subsection{Scaling potential and curvature coefficient}\label{subsec:SP}

As a central result~\cite{HPW, RPGV, EHLY, HPRW, CWESPA}, the scaling solution for the dimensionless effective potential approaches finite constants both for $\rhotil\to0$ and $\rhotil\to\infty$ (for details see sect.~\ref{sec:FESSQG})
\bel{Q5}
u(\rhotil\to0)=u_0\ ,\quad u(\rhotil\to\infty)=u_\infty\ .
\ee
For a given setting of flow equations and a given choice of the infrared regulator $R_k$ the constants $u_0$ and $u_\infty$ only depend on the numbers of effectively massless particles (scalars, fermions, gauge bosons, graviton). If for $\rhotil\to\infty$ only the massless particles of the standard model (graviton, photon, cosmon) contribute one finds
\bel{Q6}
u_\infty=\frac{5}{128\pi^2}\quad\left(-\frac{1}{128\pi^2}\right)\ ,
\ee
where the bracket also includes three generations of effectively massless neutrinos. For $\rhotil\to0$ all effectively massless particles at the UV-fixed point contribute to $u_0$. One therefore expects that $u_0$ differs from $u_\infty$. The effective scalar potential $U_k(\chi)$ of the scaling solution interpolates between $u_0k^4$ and $u_\infty k^4$. This form differs qualitatively from an almost polynomial dependence on $\chi$ that is typically found in perturbative quantum field theories for scalars. We may consider $V=u_\infty k^4$ as a type of cosmological constant induced by quantum fluctuations. Its characteristic size is given by $k$ -- the only scale present for a scaling solution.

For the curvature coefficient one finds the limiting behavior
\bel{Q7}
f(\rhotil\to0)=f_0\ ,\quad f(\rhotil\to\infty)=2\xi_\infty\rhotil\ .
\ee
This implies for large $\chi$ a $\chi$-dependent effective squared Planck mass, $F=\xi_\infty\chi^2$. The dimensionless coupling $\xi_\infty$ is the so-called non-minimal coupling of a scalar field to gravity, according to the term $-\tfrac12\xi_\infty\chi^2R$ in the effective action. Such a coupling can already be seen in perturbation theory. There may exist additional scaling solutions with constant asymptotic value $f(\rhotil\to\infty)=f_\infty$. We focus on the generic case $\xi_\infty>0$.

\begin{figure}[h]
\includegraphics{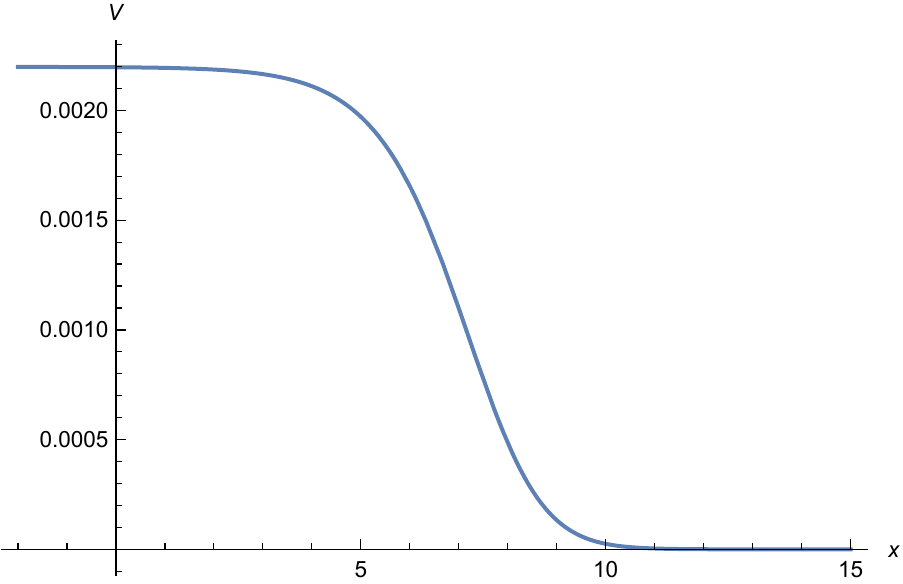}
\caption{\emph{Effective scalar potential}. We plot $\Vhat$, the potential in the Einstein frame in Planck units, as a function of the scalar field $x=\ln\rhotil=\vp/(2M)$. One observes the typical flat tail for negative and small $x$, and the exponential decrease for large $x$.}
\label{fig:1}
\end{figure}

The limiting behavior~\eqref{Q5},~\eqref{Q7} of the scaling solutions is a central result of quantum gravity formulated as a quantum field theory with associated functional flow equations. From $u(\rhotil)$ and $f(\rhotil)$ one can compute the scalar effective potential in the Einstein frame $U_E(\vp)$ (e.g. inflaton or cosmon potential) for a scalar field (see below),
\bel{15A}
U_E(\vp)=M^4\Vhat(\vp)\ ,\quad \Vhat=\frac{u(\rhotil)}{f^2(\rhotil)}\ ,\quad \vp=2M\ln\rhotil\ .
\ee
We plot this potential in Fig.~\ref{fig:1}.
It exhibits a flat tail suitable for inflation for $\vp\to-\infty$, and an exponential decrease characteristic for some form of dynamical dark energy for increasing $\vp$. The effective potential vanishes for $\vp\to\infty$.

We will argue below that this behavior is generic and rather robust. It has important consequences for cosmology. For the kinetial $K(\rhotil)$ no similarly robust results are available at present in our view, despite several encouraging first computations~\cite{HPW}.

\subsection{Relevant parameters and predictivity of\\quantum gravity}\label{subsec:RP}

For a complete quantum field theory of gravity it is sufficient that the scaling solution is approached for $k\to\infty$ or $\rhotil\to0$. Starting for arbitrarily large finite $k$ in the arbitrarily close vicinity of the scaling solution, the flow of $\Gamma_k$ towards lower $k$ may deviate from the scaling solution. This flow away from the scaling solution is typically determined in terms of a small number of \qq{relevant parameters}. To every relevant parameter one can associate a free parameter or renormalized coupling in the macroscopic quantum effective action $\Gamma_{k\to0}$. Further free parameters can arise if the scaling solution is not unique. A family of scaling solutions may be specified by continuous parameters. In any case the number of free parameters in the quantum effective action is finite. This renders our approach to quantum gravity rather predictive. If the number of free parameters is smaller than the number of renormalizable couplings in the standard model of particle physics, relations between the latter can be predicted. For cosmology one expects important restrictions on the functions $U$, $F$ and $K$, such that models with a given particle content can be tested by cosmological observations.

One of the free parameters sets the overall mass or momentum scale. Specifying only units, this is not an observable parameter. We may define a characteristic scale $k_c$ for which the solution of the flow equation starts to deviate substantially from the scaling solution due to the presence of relevant parameters. For $k>k_c$ one therefore can use the scaling solution as a good approximation. We can associate arbitrary energy units to $k_c$. Only dimensionless quantities as $\chi/k_c$ will be observable. We will discuss in sect.~\ref{sec:FSI} the attractive possibility of fundamental scale invariance~\cite{CWFSI} for which the scaling solution remains exact for all $k$. All relevant parameters vanish in this case, and the predictivity of the model is enhanced further. In this case the energy units are determined by $k$, which can again be chosen freely.

\subsection{Quantum scale invariant standard model}\label{subsec:QSSM}

In this note we mainly concentrate on the extended gravitational sector of the metric and the cosmon. These fields couple to other particles, as fermions, gauge bosons and additional scalars. The scaling solution requires that for $k\to0$ all particle physics mass scales as the Fermi scale (expectation value of the Higgs scalar) or the confinement scale in quantum chromodynamics are proportional to $\chi$. For arbitrary $k$ the electron mass takes the form $m_e=h_e(\rhotil)\chi$ and similarly the nucleon mass $m_N=h_N(\rhotil)\chi$. In the limit $k\to0$ or $\rhotil\to\infty$ the effective dimensionless coupling functions approach constants $h_{e,\infty}$ and $h_{N,\infty}$. As a consequence, the observable mass ratios electron mass over Planck mass or electron mass over nucleon mass,
\bel{Q8}
\frac{m_e}{\sqrt{F}}=\frac{h_{e,\infty}}{\sqrt{\xi_\infty}}\ ,\quad \frac{m_e}{m_N}=\frac{h_{e,\infty}}{h_{N,\infty}}\ ,
\ee
approach constants. Furthermore, dimensionless renormalizable couplings as the fine structure constant or Yukawa couplings of quarks and leptons approach constants for $\rhotil\to\infty$. For $k\to0$ these couplings do not depend on $\chi$. Even though $\chi$ typically changes in the course of the cosmological evolution one finds no time dependent fundamental couplings~\cite{CWCNC, DAPO, CHIB, UZ, DZ, DPV, CWPQT, CWTFC, MSW, VCNS, COC} for the scaling solution in this limit. Also apparent violations of the equivalence principle by a scalar-mediated fifth force are absent. The matter dominated universe shows the same observable features as for Einstein gravity, in contrast to cosmologies with a varying Planck mass and fixed particle masses~\cite{NWE, BER, FOR}.

For low enough momenta (below the effective Planck mass or some grand unification scale) the scaling solution in the limit $k\to0$ amounts to the quantum scale invariant standard model~\cite{Wetterich_1988, SZE, SHZE2, SHT}.  No intrinsic mass scale is present in the quantum effective action. All particle masses, cross sections etc. are proportional to appropriate powers of $\chi$ according to their dimension. The quantum scale invariant standard model is the basis for discussions of dark energy~\cite{Wetterich_1988, BSZ, KASH} and different versions of scale invariant inflation~\cite{GBRSZ, FHR, CPRU, FHNR, RUB, CKPR}.

The quantum scale invariant standard model remains a very good approximation for the range of $\chi$ or $\rhotil$ for which $k_c$ or $k$ are much smaller than all mass scales of the standard model. We will consider models of this type, setting $k_c$ or $k$ in the order $10^{-2}\,\text{eV}$, which is many orders of magnitude smaller than the electron mass. As a consequence, the radiation and matter dominated epochs in cosmology will be given by Einstein gravity, with possible small modifications due to the presence of the scalar field $\chi$, which may account for a small fraction of early dark energy~\cite{EDE1, EDE2, GZAV}.

The quantum scale invariant standard model does not imply that there are no running couplings. We have focused so far on vanishing momenta, as appropriate for cosmology. For scattering processes at non-zero squared momenta $q^2$ couplings as the fine structure constant $\alpha$ are running couplings. According to the scaling solution they depend on the dimensionless ratios $\rhotil$ and $q^2/\chi^2$. Quantum effects induce a running of the couplings with $q^2$ at fixed $\chi^2$, corresponding to fixed particle masses. This running follows the perturbative $\beta$-functions, with appropriate mass thresholds for the decoupling of particles~\cite{PRSM}.

\subsection{Cosmology for the scaling solution}\label{subsec:CSS}

For our discussion of cosmology we concentrate on the functions $U$, $F$ and $K$ according to the scaling solution. This covers both the setting of fundamental scale invariance and the case where the deviation from the scaling solution occurs at a scale $k_c$. In the latter case we assume that the flow stops for $k\ll k_c$. In this approximation we can use the scaling solution with $k$ identified with $k_c$. Corrections to this simplified treatment involve for $U$ or $F$ a dependence on $k/k_c$ which describes the deviation from the scaling solution. For $k\gg k_c$ this deviation vanishes. The effect of this \qq{final running} on the difference between $U(k=0)$ and $U(k=k_c)$ is typically a small constant $\sim k_c^4$, and similar for $F$ with a constant $\sim k_c^2$. We will discuss this point in sect.~\ref{sec:FSI}.

Inserting the limiting behavior of the scaling solution in the frame invariant dimensionless potential $\Vhat$~\eqref{C7} yields
\bel{Q9}
\Vhat=\frac u{f^2}\ ,\quad \Vhat(\chi\to0)=\frac{u_0}{f_0^2}\ ,\quad \Vhat(\chi\to\infty)=\frac{u_\infty k^4}{\xi_\infty^2\chi^4}\ .
\ee
This potential approaches a constant for $\chi\to0$ and vanishes $\sim\chi^{-4}$ for $\chi\to\infty$. We will find solutions of the cosmological field equations~\eqref{C5},~\eqref{C6} for which $\chi$ evolves from zero for $\eta\to-\infty$ to infinity for $\eta\to\infty$. The regime of small $\chi^2$ will be associated to inflation, while the region of large $\chi^2$ will account for dynamical dark energy. The exact vanishing of $\Vhat$ for $\chi\to\infty$ is associated to a (dynamical) solution of the cosmological constant problem.

\section{Quantum scale symmetry}\label{sec:QSS}

Quantum scale symmetry~\cite{CWQS} is a key concept for the understanding of dominant features of cosmology beyond Einstein gravity. It is directly related to the scaling solution of the flow equation and associated fixed points. At fixed points quantum scale symmetry becomes exact. For early cosmology the observed approximate scale invariance of the spectrum of primordial cosmic fluctuations can find its root in the approximate quantum scale symmetry for the vicinity of an ultraviolet (UV) fixed point for $k\to\infty$ or $\chi\to0$. Late cosmology describes the approach to an infrared (IR) fixed point for $k\to0$ or $\chi\to\infty$. Precisely at the infrared fixed point quantum scale symmetry will be an exact global symmetry of the quantum effective action. It is, however, spontaneously broken by the non-zero value of $\chi$. Any spontaneously broken exact global symmetry predicts the presence of a massless Goldstone boson -- the dilaton in our case. For finite large $\rhotil$ quantum scale symmetry is only approximate, resulting in a tiny mass for the pseudo Goldstone boson which is associated to the cosmon. Spontaneously broken approximate quantum scale symmetry gives therefore a natural reason for a very light scalar field which can provide for dynamical dark energy~\cite{Wetterich_1988}.

Quantum scale symmetry emerges as an exact global symmetry whenever the quantum effective action $\Gamma$ does not involve any intrinsic momentum or length scale. All scales are then given by fields as $\chi$. For the scaling solution of the flow equation this global symmetry is realized if for a suitable choice of fields the effective action becomes independent of $k$ for fixed fields. In our setting this typically occurs for $\rhotil\to0$ (UV-fixed point) and for $\rhotil\to\infty$ (IR-fixed point). More precisely, quantum scale symmetry is an exact global symmetry at fixed points where all scale-dependence can be absorbed in renormalized fields which transform non-trivially under scale transformations. While classical scale symmetry is broken by quantum effects leading to running dimensionless couplings, quantum scale symmetry is generated by the quantum fluctuations. It is the flow of the couplings and the associated fixed points that is responsible for this symmetry.

\subsection{Infrared fixed point}\label{subsec:IFP}

For $\rhotil\to\infty$ the quantum effective action takes the simple form
\bel{S1}
\Gamma=\int_x\sqrt{g}\bigg\{-\frac12\xi_\infty\chi^2R+u_\infty k^4+\frac12K\left(\frac{\chi^2}{k^2}\right)\partial\mo\chi\partial\mb\chi\bigg\}\ .
\ee
By rescaling $\chi$ we can set $\xi_\infty\to1$, with new kinetial $\Ktil=K/\xi_\infty$. The quantum scale transformations or dilatations act as
\bel{S2}
g\mnb\to\alpha^{-2}g\mnb\ ,\quad \chi\to\alpha\chi\ .
\ee
For $\rhotil\to\infty$ the constant potential $u_\infty k^4$ can be neglected, and quantum scale symmetry becomes exact if $\Ktil$ approaches a constant. The violations of scale symmetry close to the IR-fixed point arise for non-zero $k$ from $u_\infty k^4$ and a possible $\rhotil$-dependence of $\Ktil$. Alternatively, by use of the freedom of an overall rescaling of $\chi$ we may set $K_\infty=K(\chi\to\infty)=\pm1$. With this normalization quantum scale symmetry is realized if $\xi(\chi\to\infty)$ takes a constant value $\xi_\infty$. We will see that a negative value $K_\infty$ can be consistent with stability.

As mentioned above, the scaling solution implies that the dimensionless potential $\Vhat$ vanishes for $\chi\to\infty$. Quantum scale symmetry alone does not guarantee this behavior, since a scale invariant potential $U=\lambda\chi^4$ would lead to constant $\Vhat$. The vanishing of $\lambda$ is required by the scaling solution which only exists if $u(\rhotil\to\infty)$ approaches a constant. On rather general grounds one can show~\cite{CWIRG} that in the presence of metric fluctuations the potential $U$ cannot grow faster than $F$ for $\chi\to\infty$. This rule is obeyed for an asymptotically constant $U$, but not for $U\sim\chi^4$.

\subsection{Weyl scaling}\label{subsec:WS}

The physical implications of the effective action~\eqref{S1} are most easily understood by performing a field transformation of the metric. By the conformal transformation or Weyl scaling~\cite{HWGE, RDMPI},
\bel{S3}
g'\mnb=w^{-2}(\chi)g\mnb\ ,
\ee
the effective action~\eqref{C3} retains its form when expressed in terms of the new metric $g'\mnb$. The transformed functions are
\begin{align}
\label{S4}
F'=&w^2F\ ,\quad U'=w^4U\ ,\nn\\
K'=&w^2\bigg[K-6F\frac{\partial\ln w}{\partial\chi}\left(\frac{\partial\ln w}{\partial\chi}+\frac{\partial\ln F}{\partial\chi}\right)\bigg]\ .
\end{align}
The frame invariant combinations $\Vhat$ and $\Khat$ in eq.~\eqref{C7} remain the same when expressed in terms of $U'$, $F'$ and $K'$. The field equations~\eqref{C5},~\eqref{C6} hold for all metric frames related by an arbitrary choice of $w(\chi)$. Both conformal time $\eta$ and the combination $A$ are invariant under Weyl scalings. For many observables the independence from the choice of the metric frame has been demonstrated explicitly~\cite{FCDDGE, DTTG, CEJFC, DCECG, CCCO, PEEJF, CWEU, JIQG, JFICIM, KFCMI, CWPFF, RUCW2}. Weyl scalings change the geometry without affecting observables. Different metric frames often induce unusual pictures of cosmology~\cite{CWGE, CWUWE}.

The Einstein frame obtains for a choice $w^2=\M/F$, such that the curvature coefficient is given by the squared Planck mass, $F'=\M$. We emphasize that the Planck mass $M$ is introduced only by the variable transformation~\eqref{S3}, rather than being a parameter of the model. The scalar potential in the Einstein frame $U_E$ is directly related to the frame invariant potential $\Vhat$ by $U_E=M^4\Vhat$. For the vicinity of the IR-fixed point with $F=\xi_\infty\chi^2$ one finds for the kinetial in the Einstein frame
\bel{S5}
K_E=\frac{\M}{\chi^2}\left(\frac{K}{\xi_\infty}+6\right)\ .
\ee
The factor $\chi^{-2}$ can be absorbed by defining
\bel{S6}
\vp=4M\ln\left(\frac\chi k\right)\ ,
\ee
such that the effective action~\eqref{S1} reads in the Einstein frame,
\bel{S7}
\Gamma_E=\int_x\sqrt{g_E}\bigg\{-\frac12\M R_E+U_E(\vp)+\frac12Z(\vp)\partial\mo\vp\partial\mb\vp\bigg\}\ ,
\ee
with potential
\bel{S7A}
U_E(\vp)=\frac{u_\infty M^4}{\xi_\infty^2}\exp\left(-\frac\vp M\right)\ ,
\ee
and kinetial
\bel{S8}
Z(\vp)=\frac1{16}\left(\frac K{\xi_\infty}+6\right)\ .
\ee
We observe that the criterion of stability is $Z(\vp)\geq0$, such that $K$ can actually be negative provided $K>-6\xi_\infty$.

The potential in the Einstein frame vanishes exponentially for $\vp\to\infty$, corresponding to the vanishing of $\Vhat$ for $\chi\to\infty$. Exponential potentials are often used for models of quintessence~\cite{Wetterich_1988, RP1, VL, CLW, FEJO}. We could use a further field transform of the scalar field to bring the kinetic term to a canonical form. We will not always do so since the discussion of dynamical dark energy with a field dependent kinetial is actually quite convenient~\cite{HECW, CWCQ, CWVG, CWIQM, GKLR, KDR}. The general form of $Z$ is given by
\bel{26A}
Z=\frac{K\rhotil}{8f}+\frac38\left(\der{\ln f}{\ln\rhotil}\right)^2\ ,\quad 2\rhotil=\exp\left(\frac{\vp}{2M}\right)\ .
\ee
Stability requires $Z(\rhotil)\geq0$ for all $\rhotil$.

\subsection{Cosmon as pseudo Goldstone boson of quantum scale symmetry}\label{subsec:CPGB}

The scalar field $\sigma$ with canonical kinetic term is related to $\vp$ by
\bel{S9}
\frac{\text{d}\sigma}{\text{d}\vp}=Z^{1/2}(\vp)\ .
\ee
With this normalization the mass $m_c$ of the cosmon obeys
\begin{align}
\label{S10}
m_c^2=&\frac{\partial^2U_E}{\partial\sigma^2}=\left(1+\frac M2\frac{\partial\ln Z}{\partial\vp}\right)\frac{U_E}{Z\M}\nn\\
=&\left[1+\frac14\frac{\partial}{\partial\ln\rhotil}\ln\left(\frac K{\xi_\infty}+6\right)\right]\frac{U_E}{Z\M}\ .
\end{align}
For $\vp\to\infty$ or $\rhotil\to\infty$ the ratio $K/\xi_\infty$ becomes independent of $\rhotil$ if a fixed point is reached. The mass of the cosmon vanishes in this limit is proportional to $U_E/Z$, as expected for a pseudo Goldstone boson with explicit symmetry breaking given by $U_E$.

For $(K/\xi_\infty)(\rhotil\to\infty)=-6$ the global scale symmetry at the IR-fixed point is enhanced to a local \qq{Weyl symmetry} with spacetime-dependent parameter $\alpha(x)$ in eq.~\eqref{S2}. This includes conformal symmetry. In this limit the scalar field ceases to be a propagating degree of freedom, as seen directly from $Z(\vp\to\infty)=0$. Indeed, the metric in the Einstein frame $g'\mnb=(\xi_\infty\chi^2/M^2)g\mnb$ is invariant under local Weyl scalings. The local Weyl scaling acts as $\vp(x)\to\vp(x)+4M\ln\alpha(x)$. Local Weyl symmetry is realized if the effective action in the Einstein frame does not involve the scalar field $\vp$. If the IR-cutoff respects local Weyl symmetry, cf. refs~\cite{PERM, COPR}, this enhanced symmetry is a partial fixed point of the flow equations. (This always holds if the flow equations are compatible with the enhanced symmetry.) If this partial fixed point plays a role for the scaling solution for $\rhotil\to\infty$ one expects that $Z(\vp)$ vanishes for $\vp\to\infty$. For $Z$ vanishing slower than exponentially the cosmon mass still approaches zero for $\vp\to\infty$.

A Weyl transformation to the Einstein frame has also to be applied to fermions and other scalars in order to ensure that a standard normalization of the kinetic terms remains preserved. At the fixed point for $\vp\to\infty$ the dimensionless mass ratios or couplings become independent of $\vp$ in the Einstein frame. As a result, the cosmon can only have derivative couplings, as appropriate for a Goldstone boson. We observe that in the Einstein frame the (global) scale transformation~\eqref{S2} acts as a constant shift
\bel{S11}
\vp\to\vp+4M\ln\alpha\ ,
\ee
while the metric $g_{E\mu\nu}=g\mnb\xi_\infty\chi^2/\M$ as well as the rescaled fields for fermions, other scalars and gauge bosons are invariant. This shift symmetry implies directly the absence of non-derivative couplings of $\vp$.

\subsection{Dynamical dark energy}\label{DDE}

The potential and kinetic term of the cosmon $\vp$ are a source of dynamical dark energy, according to the field equation
\bel{S12}
H^2=\frac{1}{3\M}\left[U_E+\frac Z2\gl\partial_t\vp\gr^2+\rho_E\right]\ ,
\ee
where we have added to eq.~\eqref{C5} the contribution of the energy density in radiation and matter, given by $\rho_E$ in the Einstein frame. The scalar field evolves according to eq.~\eqref{C6}
\begin{align}
\label{S13}
\gl\partial_t^2&+3H\partial_t\gr\vp+\frac12\frac{\partial\ln Z}{\partial\vp}\gl\partial_t\vp\gr^2=-\frac1Z\frac{\partial U_E}{\partial\vp}\nn\\
=&\frac{U_E}{MZ}=\frac{M^3}{Z}\exp\left[-\frac\vp M+\ln\left(\frac{u_\infty}{\xi_\infty^2}\right)\right]\ .
\end{align}
In the limit where the term $\sim\partial\ln Z/\partial\vp$ can be neglected the scalar field $\vp$ \qq{rolls down} an exponential potential, increasing to infinity as cosmic time $t$ (or equivalently conformal time $\eta$) goes to infinity. Thus the infrared fixed point at $\vp\to\infty$ is approached asymptotically in the infinite future of the cosmic evolution. The parameters $u_\infty$, $\xi_\infty$ can be absorbed by a constant shift of $\vp$.

The homogeneous dark energy density $\rho_h$ is given by
\bel{S14}
\rho_h=U_E+\frac Z2\gl\partial_t\vp\gr^2=U_E+T_E\ ,
\ee
and the equation of state $w_h$ is defined by
\bel{S15}
w_h=\frac{T_E-U_E}{T_E+U_E}\ ,\quad T_E=\frac12(1+w_h)\rho_h\ .
\ee
Multiplying eq.~\eqref{S12} with $Z\partial_t\vp$ yields the \qq{conservation equation}
\bel{S16}
\partial_t\rho_h+6HT_E=\partial_t\rho_h+3H(1+w_h)\rho_h=0\ .
\ee
This may be compared with the conservation equation for $\rho_E$,
\bel{S17}
\partial_t\rho_E=nH\rho_E\ ,
\ee
with $n=3$ for matter domination and $n=4$ for radiation domination. For $w_h>0$ dark energy decreases faster than matter, while for $w_h<0$ the decrease of $\rho_h$ is slower than matter such that the energy density in the scalar field may finally dominate. For the matter dominated universe there exists a possible \qq{cosmic scaling solution} if $w_h=0$. In this case dark energy decreases at the same rate as matter, such that the fraction of dark energy
\bel{S18}
\Omega_h=\frac{\rho_h}{\rho_h+\rho_E}=\frac{\rho_h}{3\M H^2}\ ,
\ee
becomes a constant. For the radiation dominated universe a cosmic scaling solution with constant $\Omega_h$ is realized for $w_h=1/3$. With
\bel{S19}
y=\ln(aM)=\ln(A)\ ,
\ee
we can combine the conservation equations to
\begin{align}
\label{S20}
\partial_y\Omega_h=&-\big[3(1+w_h)+2\partial_y\ln H\big]\Omega_h\nn\\
=&\big[n-3(1+w_h)\big](1-\Omega_h)\ .
\end{align}
The last equation holds for all metric frames if we use $y=\ln A$.

The detailed dynamics of dark energy requires knowledge about the $\vp$-dependence of $Z$. We will discuss this in sect.~\ref{sec:CC}. For constant $Z<1/n$ we will indeed find cosmic scaling solutions with constant $\Omega_h=Zn$~\cite{Wetterich_1988}. They are attractors in the sense that neighboring solutions approach for increasing time these cosmic scaling solutions. Cosmic scaling solutions can give a natural explanation why the present value $\rho_h/M^4\approx10^{-120}$ is so tiny. With constant $\Omega_h$ dark energy decreases like radiation or matter, for which the small value $\rho_E/M^4\approx10^{-120}$ is naturally understood as a consequence of the huge age of the universe in Planck units. The presently observed accelerated expansion requires, however, a recent exit from such a cosmic scaling solution, for example by growing neutrino quintessence~\cite{AQCGM, CWGNCS, ABFPW, MPRC, CPW}. We will see that a cosmic scaling solution may only be reached very late in the evolution of the universe.

\subsection{Ultraviolet fixed point}\label{subsec:UVFP}

The ultraviolet fixed point corresponds to the limit $\rhotil\to0$. For fixed $k$ this is realized for $\chi\to0$, while for fixed $\chi$ it describes $k\to\infty$. For $\rhotil\to0$ the effective action according to the scaling solution is approximated by
\bel{S21}
\Gamma=\int_x\sqrt{g}\bigg\{-\frac12\gl f_0k^2+\xi_0\chi^2\gr R+\frac12 K\partial\mo\chi\partial\mb\chi+u_0k^4\bigg\}\ .
\ee
Due to the leading behavior $F=f_0k^2$, $U=u_0k^4$ the scale $k$ remains present and $\Gamma$ is not invariant under field scalings~\eqref{S2}. In contrast, neglecting the subleading term $\sim\xi_0$ and for
\bel{S22}
K=\kappa\frac{k^2}{\chi^2}\ ,
\ee
we observe a different version of quantum scale symmetry where only the scalar field $\chi$ is multiplicatively rescaled , $\chi\to\alpha\chi$, while the metric is left invariant. The leading scale symmetry violations close to this fixed point are due to $\xi_0$, as well as deviations of $K$ from the form~\eqref{S22} and corrections $\sim m_0^2\chi^2$ for $U$.

One could perform a Weyl scaling with $w^2=\chi^2/k^2$. This would replace the curvature coefficient by $F'=f_0\chi^2$ and the potential by $U'=u_0\chi^4$, while the factor $\chi^{-2}$ in $K$ would no longer be present in $K'$. In the new metric frame the effective action is invariant under the simultaneous transformations~\eqref{S2} of the metric and the scalar field. The lesson to be learned is that the quantum scale transformations at the IR- and UV-fixed points need not be the same, or the fields on which they act need not to be identical. Quantum scale symmetry at the UV-fixed point can actually also be realized if $K$ diverges for $\chi\to0$ with a power different from $\chi^{-2}$. The renormalized fields with a standard scaling behavior would then be different~\cite{CWIQM}.

At the UV-fixed point the effective action takes a particularly simple form in terms of the scalar field $\vptil$
\bel{S23}
\vptil=\sqrt{\kappa}k\ln\left(\frac\chi k\right)\ ,
\ee
namely
\bel{S24}
\Gamma=\int_x\sqrt{g}\bigg\{-\frac12f_0k^2R+u_0k^4+\frac12\partial\mo\vptil\partial\mb\vptil\bigg\}\ .
\ee
This describes a massless free scalar field with canonical kinetic term coupled to a form of Einstein gravity with a cosmological constant. The scale transformations act now as shifts in $\vptil$. From eq.~\eqref{S24} we can obtain the Einstein frame by a constant Weyl scaling with $w^2=\M/(f_0k^2)$, resulting in $U_E=u_0M^4/f_0^2$. We further transform $\vptil$ to $\vp=4M\ln(\chi/M)$, resulting in an effective action of type~\eqref{S7} with constant potential and
\bel{S25}
Z=\frac\kappa{16f_0}\ .
\ee
In the Einstein frame the solution of the field equations for $\vp\to-\infty$ is de Sitter space. (As long as corrections to quantum scale symmetry are not taken into account we can take an arbitrary constant value for $\vp$ as well.) The constant Hubble parameter reads in the Einstein frame
\bel{S26}
H_E^2=\frac{u_0M^4}{3f_0^2}\ .
\ee
De Sitter space is a good approximation for the inflationary epoch of the universe.

The primordial fluctuations of the scalar and graviton (traceless transverse tensor of the metric fluctuations) are given by the propagators of the respective fields. In turn, these propagators are determined as the inverse of the second functional derivative of the quantum effective action~\cite{CWMF}. For de Sitter space one finds an exactly scale invariant primordial fluctuation spectrum (spectral index $n_s=0$). This scale invariance is directly rooted in the quantum scale symmetry of the effective action. No intrinsic parameter with dimension of mass or length appears in the effective action once it is expressed in terms of appropriate renormalized fields. This explains why no scale appears in the fluctuation spectrum. The spectrum of the primordial cosmic fluctuations does not depend on the metric frame~\cite{CWCFVG}.

The amplitude of the graviton fluctuations obeys the frame-invariant expression\cite{CWMF, CWCFVG}
\bel{S27}
\Delta_T^2=\frac{2\cHhat^2}{\pi^2A^2}=\frac{2H_E^2}{\pi^2\M}=\frac{2U_E}{3\pi^2M^4}\ ,
\ee
where the last two equations insert the values for the Einstein frame. We may compare with the observed amplitude of the cosmic fluctuation spectrum for scalar fluctuations~\cite{PLA}
\bel{S28}
\cA=\frac{3\pi^2}{2r}\Delta_T^2=3.56\cdot10^{-8}\ ,
\ee
with tensor to scalar ratio $r<0.05$~\cite{BIKE}. This limits the value of the potential at the time when the primordial fluctuations are frozen,
\bel{46A}
r\cA=\frac{U_E}{M^4}=\Vhat=\frac{u}{f^2}\ .
\ee
Very close to the UV-fixed point this would entail the constraint $u_0/f_0^2=r\cA\lesssim2\cdot10^{-9}$. We will see in the next section that such a small value seems rather unlikely to result from a quantum gravity computation of $u_0$ and $f_0$. This is an example of typical restrictions following from quantum gravity. One concludes that the decoupling of the observed fluctuations should occur at a later time when $H_E$ is already substantially smaller than the value very close to the fixed point. This will be discussed in sect.~\ref{sec:CC}.

\section{Flow equations for quantum gravity}\label{sec:FESSQG}

This section presents the functional flow equation on which our estimates of the properties of the scaling solution and its limiting behavior for $\rhotil\to0$ and $\rhotil\to\infty$ are based. We work in second order in a derivative expansion. The scale dependent effective action $\Gamma_k$ is therefore truncated to the form~\eqref{C3}. The flow equations are evaluated for euclidean signature of the metric. Analytic continuation to Minkowski signature does not seem to pose any major problem at this level since all inverse propagators have the form $Zq^2+m^2$. We report on the flow equations for $u$ and $f$ and discuss properties of the flow equation for $K$.

\subsection{Diffeomorphism invariant flow equation for quantum gravity}\label{subsec:DIFE}

The functional flow equation for the effective average action $\Gamma_k$ and its adaption to gauge theories and gravity has been developed in ref.~\cite{Wetterich_1993, RWEAEE, MR}. We report here on the gauge invariant formulation~\cite{CWPM} which offers both technical simplifications and a direct connection to observable quantities. In this formulation the first functional derivative of $\Gamma_{k\to0}$ yields the field equations for cosmology, while the second functional derivative determines the inverse propagator. One obtains the propagator and thereby the fluctuation spectrum by inversion. In case of fundamental scale invariance the field equations and propagators can be computed for an arbitrary choice of $k$.

The exact flow equation takes the simple one loop form,
\bel{F1}
k\partial_k\Gamma_k=\frac12\text{Str}\Big\{\gl\Gamma_k^{(2)}+R_k\gr^{-1}k\partial_kR_k\Big\}-\delta_k\ .
\ee
Here $\gl\Gamma_k^{(2)}+R_k\gr^{-1}$ is the full propagator in the presence of arbitrary \qq{macroscopic fields} and the infrared regulator $R_k$. Thus both $\Gamma_k$ and $\Gamma_k^{(2)}$ are functionals of these fields and eq.~\eqref{F1} is a functional differential equation. In momentum space the supertrace $\text{Str}$ contains a momentum integral $\int_q=\int\text{d}^4q/(2\pi)^4$, a sum over different species of particles with a minus sign for fermions, as well as a trace over internal indices, including Lorentz indices $\mu$, $\nu$ or spinor indices if appropriate. The cutoff vanishes rapidly for squared momenta $q^2\gg k^2$ such that the momentum integral is ultraviolet finite due to the decay of $k\partial_kR_k$. Infrared finiteness is assured by the presence of the regulator term in the inverse propagator $\Gamma_k^{(2)}+R_k$. With a UV- and IR-finite right hand side there is no need for an additional UV-regularization. Once the flow equation is established one needs no more an explicit regularized functional integral respecting the symmetries. The microphysics is encoded in the \qq{initial conditions} of $\Gamma_k$ for $k\to\infty$. This implicit \qq{ERGE regularization} constitutes an important advantage for theories for which no explicit gauge invariant regularization is known, as in the case of quantum gravity. Finally, $\delta_k$ is a \qq{measure factor} which accounts for the redundant formulation in case of local gauge theories.

For the gauge invariant formulation of the flow equation the first term on the r.h.s. involves a projection on the physical fluctuations. This is effectively achieved by a suitable \qq{physical gauge fixing}. In this formulation the conceptual structure of the first term is completely analogous to simpler theories for scalars and fermions without local gauge invariance. The \qq{measure factor} is given by a simple functional (typically a derivative of a regularized determinant) that does not depend on $\Gamma_k$. Ghosts need not to be introduced for this purpose since the Faddeev-Popov determinant can be regularized directly. We will not describe here all computational steps leading to the flow equations for $u$, $f$ and $K$. We only present the main lines and the results which have a simple intuitive form.

\subsection{Flow equation for effective potential}\label{subsec:FEEP}

We first evaluate the flow equation for constant scalar fields and a constant metric. Since all derivatives vanish this projects $\Gamma_k$ on the effective potential for the scalar fields, multiplied by $\sqrt{g}$, i.e. $\Gamma_k=\int_x\sqrt{g}U_k$. Correspondingly, $\Gamma_k^{(2)}$ has to be evaluated for constant macroscopic fields. In the presence of these fields one typically finds momentum independent contributions to $\Gamma_k^{(2)}$. These field dependent \qq{mass terms} are functions of the constant scalar fields.

The flow equation for the effective scalar potential $U_k$ can be written in an intuitive form~\cite{PRWY, CWESPA}
\begin{align}
\label{F2}
k\partial_kU_k=&\Pit_U=\Pit\subt{grav}+\Pit_s+\Pit\subt{gauge}+\Pit_f\nn\\
=&\frac{k^4}{32\pi^2}\gl2\Nbar_g+\Nbar_S+2\Nbar_V-2\Nbar_F\gr=4k^4c_U\ .
\end{align}
Different parts arise from fluctuations of different fields, with $\Nbar_j$ the effective numbers of particle species as described below.

\zwisch{Metric fluctuations}

The first contribution arises from the metric fluctuations
\bel{F3}
\Pit\subt{grav}=\frac{k^4}{24\pi^2}\left(1-\frac{\eta_g}{8}\right)\left(\frac5{1-v}+\frac1{1-v/4}\right)-\frac{k^4}{8\pi^2}\ .
\ee
It depends on the dimensionless ratio
\bel{F4}
v=\frac{2U}{Fk^2}=\frac{2u}{f}\ .
\ee
Here the first term in eq.~\eqref{F3} reflects the five degrees of freedom of the traceless tensor fluctuations (graviton fluctuations) whose propagator involves an effective mass term $-2U/F$. The second term is due to the physical scalar degree in the metric fluctuation with effective mass term $-U/(2F)$. (Here physical fluctuations are defined in contrast to the pure gauge fluctuations. This does not mean that the physical scalar metric fluctuation, which accounts for Newton's potential, is propagating as a particle. The particle degrees of freedom are only two polarizations of the graviton.) Finally, the constant last term reflects the metric contribution to the measure factor $\delta_k$.

For the precise form of the flow equations we follow ref.~\cite{PRWY}, for early investigations see ref.~\cite{NAPE, DELP, EIPAU,  LPSW}. We have taken a particular form of the infrared cutoff function, namely a Litim-type regulator~\cite{LIT} $R_k\sim\gl k^2-q^2\gr\theta\gl k^2-q^2\gr$. This replaces an inverse propagator $q^2$ by $k^2$ if $q^2<k^2$, and does not change the propagator for $q^2>k^2$, leading to $k\partial_kR_k=0$ for $q^2>k^2$. For this regulator a mass term $m^2$ in the inverse propagator $\sim\gl q^2+m^2\gr$ generates a \qq{threshold function}
\bel{F5}
s(\mtil^2)=\gl1+\tilde m^2\gr^{-1}\ ,\quad \mtil^2=\frac{m^2}{k^2}\ ,
\ee
which multiplies the contribution of the massive particle. This threshold function leads to an automatic suppression of the contribution from particles with $m^2>k^2$, such that the flow equation incorporates naturally the decoupling of heavy particles. This contrasts to many other regularization schemes as dimensional regularization. For other choices of the regulator $R_k$ the precise form of the threshold function will differ, but the qualitative decoupling behavior remains the same.

We also observe a pole in the threshold function for $\mtil^2\to-1$. This can be related to convexity properties of the effective potential~\cite{TETW}. For the graviton contribution the factor $(1-v)^{-1}$ reflects this threshold function, with $\mtil^2=-v$, and similar for the scalar metric fluctuation with $\mtil^2=-v/4$. We note that the effective mass term $\mtil^2$ for the metric fluctuations is negative for positive $u$ and $f$. Values of $v$ close to one can substantially enhance the impact of the graviton fluctuations. They dominate for the range of positive $v$.

Finally the quantity $\eta_g$,
\bel{F6}
\eta_g=-k\partial_k\ln f=2-\frac{k\partial_k F}{F}\ ,
\ee
reflects that the regulator for the metric fluctuations is taken proportional to $F$. Thus $\eta_g$ vanishes for constant $f$ and equals two if $F$ is independent of $k$. In the limit $|v|\ll1$ one finds for constant $F$
\bel{F6A}
\Pit\subt{grav}=\frac{k^4}{16\pi^2}\ ,\quad \Nbar_g=1\ .
\ee
This corresponds to the contribution of the two propagating degrees of freedom of the graviton. In our setting for cosmology this limit applies for $k^2\ll\chi^2$.

\zwisch{Scalar fluctuations}

The contribution from scalar fluctuations is the same as for models without gravity
\bel{F7}
\Pit_s=\frac{k^4}{32\pi^2}\sum_A\left(1-\frac{\eta_A}{6}\right)\left(1+\mtil_A^2\right)^{-1}=\frac{\Nbar_Sk^4}{32\pi^2}\ ,
\ee
where the sum runs over $N_S$ scalar fields. The index $A$ labels the eigenvalues $m_A^2$ of the renormalized scalar mass matrix $M^2$,
\bel{F8}
M^2_{ab}=\gl Z_aZ_b\gr^{-1/2}\frac{\partial^2U}{\partial\phi_a\partial\phi_b}\ ,\quad \mtil_A^2=\frac{m_A^2}{k^2}\ .
\ee
Here $Z_a$ is the kinetial of the scalar field $\phi_a$, $a=1\dots N_S$, $\eta_a=-k\partial_k\ln Z_a$. The anomalous dimension $\eta_A$ arises from $Z_A$ multiplying $R_k$ and is identified with some suitable $\eta_a$. It is typically a small quantity and can be neglected. We identify $\Nbar_S$ in eq.~\eqref{F2},~\eqref{F7} with the effective number of real scalar fields. For $\eta_A=0$ it coincides with the number of effectively massless scalars for which $\mtil_A^2\ll1$. Since $\mtil_A^2$ depends on the values of the constant macroscopic scalar fields the effective number $\Nbar_S$ varies in different regions of field space and for different $k$. Inbetween mass thresholds one finds, however, an (almost) constant value of $\Nbar_S$. The overall picture is simple: every effectively massless scalar contributes to $\Pit_s$ a term $k^4/(32\pi^2)$. For massless scalars the only mass scale is given by $k$, such that the factor $k^4$ is dictated by the dimension of the scalar potential $U_k$.

The flow equation~\eqref{F2},~\eqref{F3},~\eqref{F7} is derived in the truncation of variable gravity~\eqref{C3} which includes terms with up to two derivatives. Within this truncation we have omitted a subleading term. For $\partial U/\partial\phi=0$ the scalar degree of freedom in the metric fluctuations mixes with the other scalars $\phi_a$. The resulting correction term~\cite{PRWY} $\sim\gl\partial U/\partial\vp\gr^2$ vanishes at the minimum of $U$ and can be neglected for sufficiently flat $U$.

We evaluate $U(\chi)$ at the partial minimum with respect to other additional scalar fields as the Higgs doublet. At the partial minimum these additional scalar fields do not mix with $\chi$ through the mass matrix. They also do not mix with the metric fluctuations. Then the additional scalars decouple in a range of $k$ smaller than their masses. These properties single out the definition of $\chi$ at the partial minimum with respect to the other scalars. For a different choice the flow equations would be more complicated and do not feature the effective decoupling. (Our flow equations concern the effective potential at zero temperature. If fields are displaced from their minimum in vacuum due to temperature effects the situation gets more complex.) For $k$ smaller than the mass of the lightest additional scalar only the fluctuations of $\chi$ contribute to $\Pit_s$, with
\bel{F9}
\mtil^2=\frac{\partial u}{\partial\rhotil}+2\frac{\partial^2u}{\partial\rhotil^2}\ .
\ee
In the region where $u(\rhotil)$ is flat one has approximately $\mtil^2=0$ and therefore $\Nbar_S=1-\eta_s/6$. The mixing with the metric fluctuations due to $(\partial U/\partial\chi)^2$ can be neglected in these regions.

\zwisch{Fermion and gauge boson fluctuations}

The contributions from fermion fluctuations is even simpler. For effectively massless fermions $\Nbar_F$ counts the number of Weyl-fermions or equivalently Majorana fermions. For example, in the region of $k\gg m_e$ the electron fluctuations contribute $\Nbar_{F,e}=2$ as appropriate for a Dirac fermion which is constituted of two Majorana fermions or Weyl fermions. In general, we consider $N_F$ Majorana fermions with masses $m_f^2$ and $\mtil_f^2=m_f^2/k^2$. Neglecting possible small anomalous dimensions for the fermionic kinetic terms they contribute
\bel{F10}
\Pit_f=-\frac{\Nbar_Fk^4}{16\pi^2}\ ,\quad \Nbar_F=\sum_{f=1}^{N_f}\gl1+\mtil_f^2\gr^{-1}\ .
\ee
We observe again the decoupling of fermions once $m_f^2\ll k^2$. For the scaling solution the fermion masses are given by effective dimensionless Yukawa couplings $h_f$,
\bel{F11}
m_f=h_f\chi\ ,\quad \mtil_f^2=2h_f^2\rhotil\ .
\ee

For the contribution $\Pit\subt{gauge}$ from gauge bosons $\Nbar_V$ counts the number of effectively massless gauge bosons. For each massless gauge boson the physical fluctuations are the transversal fluctuations and contribute a factor three.  The measure term subtracts one, resulting in the expression $2\Nbar_V$ in eq.~\eqref{F2}, which reflect the two polarizations of a propagating massless vector field. Gauge bosons can acquire masses $m_v$ through the Higgs mechanism, $\mtil_v^2=m_v^2/k^2$, such that $\Nbar_V$ is approximated by
\bel{F12}
2\Nbar_V=\sum_{v=1}^{N_V}\Big[3\gl1+\mtil_v^2\gr^{-1}-1\Big]\ .
\ee
The three massive gauge boson degrees of freedom decouple for $\mtil_v^2\gg1$. What remains is the measure factor which does not involve $\mtil_v^2$. This measure factor cancels precisely the contribution of the massless Goldstone boson in $\Nbar_S$. For each massive gauge boson there is one massless Goldstone boson that transmutes to the longitudinal massive gauge boson. As a result, only the three degrees of freedom of the massive gauge boson and the massive non-Goldstone scalar modes (\qq{radial modes}) contribute to the flow. They decouple once all mass terms exceed $k^2$. 

For the quantum scale invariant standard model the masses of the W- and Z-bosons are proportional to the Fermi scale $\vp_0$, which in turn is proportional to $\chi$. This results in $\mtil_v^2=c_v\rhotil$, with very small $c_v\sim g^2\vp_0^2/\chi^2$ involving the gauge coupling $g$ and the tiny ratio $\vp_0^2/\chi^2$. This ensures that the W- and Z-bosons decouple only once $k$ gets smaller than their mass.

\zwisch{Robustness of flow equation}

In summary, a rough approximation to the flow of $U_k$ simply counts the degrees of freedom for massless particles, consisting of $\Nbar_g$ gravitons, $\Nbar_S$ scalars, $\Nbar_F$ fermions and $\Nbar_V$ gauge bosons, as relevant for a given range of $k$ or $\rhotil$. In view of this very simple structure, where only the physical propagating modes contribute in the range where their masses are smaller than $k$, the flow equation for $U$ seems to be rather robust.

In the gauge invariant formulation of the flow equation propagators and vertices obtain by taking appropriate derivatives of $\Gamma_k$. From the potential $U_k$ we obtain the mass matrix by taking two derivatives
\bel{F12A}
\Mbar^2_{ab}=\frac{\partial^2 U}{\partial\vp_a\partial\vp_b}\ .
\ee
Correspondingly, the flow equation for $\Mbar^2_{ab}$ is found by taking two derivatives of the flow generator
\bel{F12B}
k\partial_k\Mbar^2_{ab}=\frac{\partial^2\Pit_U}{\partial\vp_a\partial\vp_b}\ .
\ee
This generalizes to the flow of vertices, as quartic scalar couplings which involve four derivatives of $U$. Omitting the contributions of metric fluctuations, which is suppressed for $k^2\ll F$, this procedure reproduces the perturbative $\beta$-functions for the running quartic couplings in one loop order, plus part of the higher loop contributions. Exact two-loop $\beta$-functions require an extended truncation~\cite{PACW}. It is interesting to note that it is precisely the threshold functions for the decoupling of massive particles which are responsible for the running quartic couplings. The field-dependence of $k\partial_kU_k$ arises only from these threshold functions which induce a field-dependence of $\Nbar_S$, $\Nbar_V$, $\Nbar_F$. The one-loop $\beta$-functions are universal in the sense that they do not depend on the choice of the IR-cutoff $R_k$. The fact that perturbative $\beta$-functions obtain in a straightforward way from the flow equation~\eqref{F2} enhances our confidence in the validity of this approach.

\subsection{Flow equation for curvature coefficient}\label{subsec:FECC}

For extracting the flow equation for $F$ we continue to take constant scalar fields. In contrast, we consider a metric $g\mnb(x)$ different from the constant metric for flat space. We choose this metric such that the associated curvature scalar $R$ is small. By evaluating the difference of the flow of $\Gamma_k$ as compared to the flow in flat space one extracts the flow of $-\frac12\int_x\sqrt{g}FR$ and therefore $F$ or $f$. For the evaluation of the flow generator one may use heat kernel methods for general metrics or specialize to particular metric configurations as the ones for spheres.

Following ref.~\cite{CWMY} the gauge invariant flow equation leads to
\begin{align}\label{F13}
k\partial_kF=&2k^2c_F=2k^2\gl c^{(\text{grav})}_F+c^{(S)}_F+c^{(F)}_F+c^{(V)}_F\gr\nn\\
=&2k^2c^{(\text{grav})}_F+\frac{k^2}{48\pi^2}\gl-\Nbar_S-\Nbar_F+4\Nbar_V'\gr\ .
\end{align}
The contribution from metric fluctuations is given by
\bel{F14}
c_F^{(\text{grav})}=\frac{25(1-\eta_g/6)}{64\pi^2(1-v)}-\frac{(1-11\eta_g/64)}{72\pi^2(1-v/4)}+\frac{17}{192\pi^2}\ .
\ee
The first term arises from the traceless tensor or graviton fluctuations, the second approximates the contribution of the scalar metric fluctuations and the last term reflects the measure contribution for the metric sector. We observe that the graviton fluctuations dominate over the scalar metric fluctuations by a large factor (unless $v$ takes very large negative values). We have again simplified the scalar sector by omitting the mixing of the scalar metric fluctuations with additional scalar fields. (For an explicit expression for this small correction see ref.~\cite{CWMY}.)

The contribution from scalar fluctuations is given by
\bel{F15}
c_F^{(S)}=-\frac{\Nbar_S}{96\pi^2}+c_F^{(\xi)}\ ,
\ee
with effective number of massless scalars $\Nbar_S$ given by eq.~\eqref{F7}. The second term arises from the field-dependence of the curvature coefficient. For a single scalar field with inverse propagator (neglecting mixing with the scalar metric fluctuation)
\begin{align}\label{F16}
G^{-1}=&Zq^2+\frac{\partial^2U}{\partial\chi^2}-\frac12\frac{\partial^2F}{\partial\chi^2}R\nn\\
=&Z(q^2+m^2-\tilde\xi R)\ ,
\end{align}
where $\tilde\xi=(\partial^2F/\partial\chi^2)/(2Z)$ one finds~\cite{CWMY}
\bel{F17}
c_F^{(\xi)}=-\frac{\tilde\xi}{32\pi^2(1+\mtil^2)^2}\ .
\ee
In turn, the flow equation for $\tilde\xi$ obtains by taking the second $\chi$-derivative of eq.~\eqref{F13}, see ref.~\cite{NAPE} for an early computation. Off-diagonal kinetic terms mixing the scalar metric fluctuations with the fluctuations of the additional scalar render the situation more complex. We will in the following omit this contribution, keeping in mind that a better understanding is needed.

The fermion contribution obtains as
\bel{F18}
c_F^{(F)}=-\frac{\Nbar_F}{96\pi^2}\ ,
\ee
with $\Nbar_F$ given by eq.~\eqref{F10}. For the contribution of gauge boson fluctuations one finds
\bel{F19}
c_F^{(V)}=\frac{4\Nbar_V'}{96\pi^2}\ ,\quad 4\Nbar_V'=\sum_v\big[3(1+\mtil_v^2)^{-1}+1\big]\ .
\ee
The last constant term is the measure contribution from the gauge sector for which we note the opposite sign as for the contribution to the flow of the effective potential. This contribution cancels the contribution of the Goldstone boson from $c_F^{(S)}\sim-\Nbar_S$. Again, massive gauge bosons decouple in the limit $\mtil_v^2\gg1$, with only the three massive degrees of freedom contributing.

The sign of $c_F^{(\text{grav})}$ and $c_F^{(V)}$ is positive, while contributions from scalar and fermion fluctuations have the opposite sign with negative $c_F^{(S)}$ and $c_F^{(F)}$. An overall positive sign of $c_F$ restricts the number of scalars and fermions.

\subsection{Flow equation for kinetial}\label{subsec:FEK}

The flow equation for the kinetial $K$ is not yet known reliably. A reliable computation needs to reproduce the property that for the scalar-gravity system one has an enhanced local Weyl symmetry in the limit $\chi\to\infty$, $K/\xi_\infty\to-6$. After a Weyl scaling to the Einstein frame the scalar $\vp$ appears no longer in the effective action if $Z=0$ and $\partial U/\partial\vp=0$. This reflects the enhanced local symmetry which transmutes the scalar degree of freedom to a pure gauge degree of freedom of this enhanced symmetry. For this limit the Einstein frame can be viewed as removing the gauge degree of freedom of local Weyl symmetry which no longer couples to the physical sector.

As a result of this enhanced symmetry the flow of $Z$ should either vanish for $Z=0$ and $\partial U/\partial\vp=0$ or it should diverge. In both cases the value $Z=0$ cannot be reached. For the case of vanishing flow the fixed point can be approached asymptotically and one expects for $\partial U/\partial\vp\to0$ a flow equation of the type
\bel{F20}
k\partial_kZ=-\eta_ZZ\ ,
\ee
where the anomalous dimension $\eta_Z$ can depend on other couplings and may vanish. In turn, this translates to
\bel{F21}
k\partial_k\left(\frac K\xi\right)\gl\chi\to\infty\gr=-\eta_Z\left(\frac K\xi+6\right)\ ,
\ee
with
\bel{F22}
\xi=\frac12\frac{\partial F}{\partial\rho}=\frac1{2\chi}\frac{\partial F}{\partial\chi}\ .
\ee
It is a somewhat involved task to reproduce the property~\eqref{F21} from the flow equations for $K$ and $F$. Perturbative one-loop results for these flow equations can be inferred from ref.~\cite{STKA}.

The flow equation is formulated at fixed $g\mnb$ and $\chi$ and not for fixed fields $g'\mnb$ and $\vp$ in the Einstein frame. The nonlinear field transformation from $\chi$ to $\vp$ and from $g\mnb$ to $g'\mnb$ depends on $k$. The transformation of the flow equation under a change of field variables is well known and results in an additional term~\cite{CWFT, GIWE, PAWR, FLOWE}
\begin{align}
\label{F23}
k\partial_k\Gamma|_{g'\mnb,\vp}=&k\partial_k\Gamma|_{g\mnb,\chi}-\int_x\frac{\partial\Gamma}{\partial g'\mnb}k\partial_kg'\mnb|_{g\mnb,\chi}\nn\\
&-\int_x\frac{\partial\Gamma}{\partial\vp(x)}k\partial_k\vp(x)|_{g\mnb,\chi}\ .
\end{align}
Furthermore, the second functional derivative $\Gamma_k^{(2)}$ has to be translated to functional derivatives with respect to $g'\mnb$. (There is a possibility to formulate the IR-cutoff term in terms of $g'\mnb$ and $\vp$, which yields a formulation where the second functional derivative with respect to these fields appears in the flow equation. Of course, the same cutoff function has to be used for the computation of $k\partial_kU$, $k\partial_kF$ and $k\partial_kK$.) For the Weyl scaling to the Einstein frame, $g'\mnb=(F/M^2)g\mnb$, eq.~\eqref{F23} reads
\begin{align}\label{F24}
k\partial_k\Gamma|_{g'\mnb,\vp}=&k\partial_k\Gamma|_{g\mnb,\chi}+\frac{k^2c_F}{M^2}\int_x\sqrt{g'}\Ttil_E\mno g'\mnb\nn\\
&+4M\int_x\frac{\partial\Gamma}{\partial\vp(x)}\ ,
\end{align}
where the second term involves the trace of the energy momentum tensor $\Ttil_E\mno$ in the Einstein frame, including here a \qq{gravitational part} according to $\partial\Gamma/\partial g'\mnb=-\gl\sqrt{g'}/2\gr\Ttil_E\mno$. For solutions of the field equations the additional terms vanish.

Eq.~\eqref{F24} may permit to compute the flow of the wave function in the Einstein frame and to devise a form of the cutoff function $R_k$ that is compatible with the enhanced local Weyl symmetry. In the Einstein frame one may expect that for small $k^2/M^2$ the metric fluctuations effectively decouple for the flow of $Z$. This short discussion demonstrates the work that needs to be done.

\section{Scaling solution}\label{sec:SS2}

The scaling solution plays a central role for our discussion of cosmology. In this section we therefore investigate the scaling solution for the dimensionless functions $u(\rhotil)$ and $f(\rhotil)$ in some detail. In particular, we discuss the robustness of the important limiting behavior for $\rhotil\to0$ and $\rhotil\to\infty$. The functions $u(\rhotil)$ and $f(\rhotil)$ are sufficient in order to determine the frame-invariant potential $\Vhat(\rhotil)$ for the scaling solution. For the second frame invariant function $\Khat(\rhotil)$ further computation is required.

\subsection{Differential equations for scaling solutions}

A scaling solution requires that $u$ and $f$ are only functions of $\rhotil=\chi^2/(2k^2)$, solving the flow equation at fixed $\rhotil$. The flow equations~\eqref{F2} and~\eqref{F13} are evaluated at fixed $\rho=\chi^2/2$
\bel{F25}
k\partial_k u|_{\rho}=-4u+4c_U\ ,\quad k\partial_k f=-2f+2c_F\ .
\ee
The flow equations at fixed $\rhotil$ obtain by a variable change
\bel{F26}
k\partial_k u|_{\rhotil}=k\partial_k u|_{\rho}-\frac{\partial u}{\partial\rhotil}k\partial_k\rhotil|_{\rho}
\ee
where
\bel{F27}
k\partial_k\rhotil|\rho=-2\rhotil\ .
\ee
and similar for $f$. In terms of the dimensionless scalar fields eq.~\eqref{F25} transforms to
\begin{align}\label{F28}
\gl k\partial_k-2\rhotil\partial_{\rhotil}\gr u=&4(c_U-u)\ ,\nn\\
\gl k\partial_k-2\rhotil\partial_{\rhotil}\gr f=&2(c_F-f)\ .
\end{align}
In our truncation the quantities $c_U$ and $c_F$ are functions of $\rhotil$, involving $u$, $f$ and derivatives thereof. Eq.~\eqref{F28} constitutes a closed system of differential equations.

The scaling solution solves eq.~\eqref{F28} with $k\partial_k u|_{\rhotil}=0$, $k\partial_k f|_{\rhotil}=0$. We infer the differential equations that scaling solutions have to obey,
\bel{F29}
\rhotil\partial_{\rhotil} u=2(u-c_U)\ ,\quad \rhotil\partial_{\rhotil} f=f-c_F\ .
\ee
These are central equations for this work. The properties of the scaling functions $u(\rhotil)$ and $f(\rhotil)$, and in particular their behavior in the limits $\rhotil\to0$ and $\rhotil\to\infty$, follow from the solutions of these differential equations. As a boundary condition we require that for $\rhotil\to0$ both $u$ and $f$ reach finite values
\bel{F30}
u(\rhotil=0)=u_0=c_U(\rhotil=0)\ ,\quad f(\rhotil=0)=f_0=c_F(\rhotil=0)\ .
\ee
In turn, this requires finite values of $c_U$ and $c_F$ for $\rhotil\to0$. Together with the requirement that $u$ and $f$ (and therefore also $c_U$ and $c_F$) should remain finite for any finite value of $\rhotil$ these conditions severely restrict the possible scaling solutions.

Keeping in mind the corrections discussed above we approximate the flow generators for $u$ by
\bel{F31}
c_U=\frac{1}{96\pi^2}\gl1-\frac14\rhotil\partial_{\rhotil}\ln f\gr\left(\frac{5}{1-v}+\frac{1}{1-v/4}\right)+\frac{\cN_U}{128\pi^2}\ ,
\ee
with
\bel{F32}
\cN_U=\Nbar_S+2\Nbar_V-2\Nbar_F-4\ .
\ee
The generator $c_F$ will be approximated by
\bel{F33}
c_F=\frac{(1-\frac13\rhotil\partial_{\rhotil}\ln f)}{64\pi^2}\left(\frac{25}{1-v}-\frac{8}{9(1-v/4)}\right)+\frac{\cN_F}{96\pi^2}\ ,
\ee
where
\bel{F34}
\cN_F=-\Nbar_S-\Nbar_F+4\Nbar_V'+\frac{17}{2}\ .
\ee
Here we employ for the scaling solution $\eta_g=2\rhotil\partial_{\rhotil}\ln f$ and we simplify its slightly different role for the graviton and scalar metric fluctuations. We have omitted the term $c_F^{(\xi)}$ in eq.~\eqref{F15}. The main corrections are presumably due to the omission of mixing between the scalar metric fluctuations and additional scalars. Since the contribution of the scalar metric fluctuations is substantially smaller than the one from the graviton fluctuations (\qq{graviton dominance}) the main characteristics should be well described by the approximation~\eqref{F31}-~\eqref{F34}.

The range of validity of the flow equations is restricted to $v<1$. For $v=1$ the graviton propagator is divergent even in presence of the IR-cutoff. One finds that $v=1$ constitutes a barrier in the flow that is not crossed~\cite{CWIRG}. This is related to general convexity properties of the scale-dependent effective action~\cite{TETW}.

\subsection{Limiting behavior of scaling solutions}

For large $\rhotil\to\infty$ one finds a simple solution
\bel{F35}
f(\rhotil)=2\xi_\infty\rhotil\ ,\quad F=\xi_\infty\chi^2\ .
\ee
Indeed, for finite $c_F$ a term $2\xi_\infty\rhotil$ dominates the r.h.s. of eq.~\eqref{F29} for $\xi_\infty\neq0$. A similar solution $u\sim\rhotil^2$ is not possible since it would lead to divergent $v$ for positive $u$, or to negative divergent $u$ which is forbidden by convexity properties in the scalar sector. What remains is a constant value
\bel{F36}
u(\rhotil\to\infty)=u_\infty=\frac1{128\pi^2}\gl2+\Nbar_S+2\Nbar_V-2\Nbar_F\gr\ .
\ee
Here we note that for $f\sim\rhotil$, $u\to u_\infty$ one has $v\to0$, $\eta_g\to2$. At this point we have a whole family of possible scaling solutions parameterized by $\xi_\infty$. Not all of them may correspond to true scaling solutions that remain valid for the whole range $0\leq\rhotil<\infty$. For $\xi_\infty\neq0$ we can expand the flow equation in inverse powers of $\rhotil$, with fixed coefficients for given $\xi_\infty$~\cite{HPRW}. The case $\xi_\infty=0$ is special. It corresponds to the constant scaling solution (generalized Reuter fixed point) for which $u$ and $f$ are independent of $\rhotil$ and take the same value as for $\rhotil=0$.

In the limit $\rhotil\to0$ both $u$ and $f$ approach constants
\begin{align}\label{F37}
u(\rhotil\to0)=&u_0=c_U(0)\nn\\
=&\frac{1}{128\pi^2}\bigg[\frac43\left(\frac{5}{1-v_0}+\frac{1}{1-v_0/4}\right)\nn\\
&\quad\quad\quad-4+\Nbar_S+2\Nbar_V-2\Nbar_F\bigg]\ ,
\end{align}
and
\begin{align}\label{F38}
f(\rhotil\to0)=&f_0=c_F(0)\nn\\
=&\frac1{96\pi^2}\bigg[\frac32\left(\frac{25}{1-v_0}-\frac{8}{9(1-v_0/4)}\right)\nn\\
&\quad\quad\quad+\frac{17}{2}-\Nbar_S-\Nbar_F+4\Nbar_V'\bigg]\ .
\end{align}
Here we employ $v_0=u_0/(2f_0)$ and $\eta_g=0$. Inserting this value eqs.~\eqref{F37},~\eqref{F38} are two coupled non-linear equations for $u_0$ and $f_0$. A discussion of the possible solutions with $f_0>0$ in dependence on the numbers of effectively massless particles can be found in ref.~\cite{CWMY}, or, for somewhat different flow equations, in ref.~\cite{RPDP, DEP, CLPM, AE1}.

\subsection{Scaling solution for potential}

For scaling solutions with $\rhotil$-dependent masses or $\xi_\infty\neq0$ the effective numbers of particles are different for $\rhotil\to0$ and $\rhotil\to\infty$. As $\rhotil$ increases, more and more particles decouple from the flow since their masses become larger than $k$. We may write
\bel{88A}
c_U=\sum_jc_U^{(j)}+c_U\supt{grav}-\frac{N_V^{(0)}}{128\pi^2}\ ,
\ee
with
\bel{88B}
c_U^{(j)}=\frac{N_j^{(u)}}{128\pi^2(1+\tau_j\rhotil)}\ ,
\ee
the contribution of particles with mass given by $m_j^2=\tau_j\chi^2/2$, $\mtil_j^2=\tau_j\rhotil$. Here $N_j^{(u)}=N_{S,j}+3N_{V,j}-2N_{F,j}$ involves the appropriate combination of scalars, gauge bosons and Majorana fermions with mass $m_j$. (This effective number may include contributions from the anomalous dimension.) The number $N_V^{(0)}$ denotes the number of massless gauge bosons and the last term in eq.~\eqref{88A} arises from the measure term for the massless gauge bosons. For the massless gauge bosons this subtracts one unit from $3N_{V,j}$, such that only $2N_V^{(0)}$ massless degrees of freedom contribute. The contributions of the $N_V-N_V^{(0)}$ Goldstone bosons are approximated here as massless even away from the potential minimum. As we have discussed before, their contribution is canceled by the measure terms for the $N_V-N_V^{(0)}$ massive gauge bosons. In consequence, neither the Goldstone boson fluctuations not the measure terms for the massive gauge bosons appear in $\sum_jc_U^{(j)}$ in eq.~\eqref{88A}.

The scaling solution of the differential equation~\eqref{F29} yields
\bel{88C}
u=\sum_ju_j+u\subt{grav}\ ,
\ee
where
\bel{88CA}
u_j=\frac{N_j^{(u)}}{128\pi^2}t_u\gl\tau_j\rhotil\gr
\ee
involves the threshold function $t_u\gl\tau_j\rhotil\gr$. This threshold function,
\bel{88D}
t_u\gl\tau_j\rhotil\gr=1-2\tau_j\rhotil-2\gl\tau_j\rhotil\gr^2\ln\left(\frac{\tau_j\rhotil}{1+\tau_j\rhotil}\right)\ ,
\ee
interpolates between the limits
\bel{88E}
t_u\gl\tau_j\rhotil\ll1\gr=1-2\tau_j\rhotil\ ,\quad t_u\gl\tau_j\rhotil\gg1\gr=\frac2{3\tau_j\rhotil}\ ,
\ee
and obeys
\bel{94A}
y\der{t_u(y)}{y}=2t_u(y)-\frac2{1+y}\ .
\ee

The metric contribution obeys
\begin{align}
\label{88F}
\gl\rhotil&\partial_{\rhotil}-2\gr u\subt{grav}=-2c_{U,\text{grav}}\ ,\nn\\
c_{U,\text{grav}}=&\frac1{96\pi^2}\left(1-\frac14\rhotil\partial_{\rhotil}f\right)\left(\frac5{1-v}+\frac1{1-v/4}\right)-\frac1{32\pi^2}\ .
\end{align}
For large $f\approx2\xi\rhotil$ one has $v\to0$ and $u\subt{grav}$ approaches a constant
\bel{88G}
u\subt{grav}=\frac1{64\pi^2}+O\left(\frac1f\right)\ .
\ee
Together with $N_V=1$ for the photon and $N_S=1$ for the scalar field this yields eq.~\eqref{Q6}.

\subsection{Neutrinos, standard model and grand unification}

Neutrinos are much lighter than the other fermions by virtue of the see-saw mechanism~\cite{MIN, YAN, GRS, MACW, LSW} and play an interesting role. For $k$ smaller than the neutrino masses only the metric fluctuations, the photon and the cosmon contribute to the flow leading to $2+\Nbar_S+2\Nbar_V=5$, and therefore positive $u_0$. Once $k$ exceeds the heaviest of the neutrino masses one finds $2+\Nbar_S+2\Nbar_V-2\Nbar_F=-1$, and therefore negative $c_U$. If we simplify to three equal neutrino masses, $m_\nu=h_\nu M$ in the Einstein frame, the scaling equation for $u$ in the region where the \qq{neutrino threshold} is crossed reads ($\mtil_\nu^2=2h_\nu^2\rhotil$, $\tau_\nu=2h_\nu^2$)
\bel{F39}
\rhotil\partial_{\rhotil}u=2u-\frac1{64\pi^2}\left(5-\frac6{1+2h_\nu^2\rhotil}\right)\ .
\ee
Starting from positive $u\approx u_0$ for $\rhotil\gg h_\nu^{-2}$, and decreasing $\rhotil$, the positive r.h.s. of eq.~\eqref{F39} drives $u$ to smaller values. For $h_\nu^2\rhotil\gg1$ this results in
\bel{F40}
u=\frac{5}{128\pi^2}-\frac{1}{64\pi^2h_\nu^2\rhotil}\ ,
\ee
while for $h_\nu^2\rhotil\ll1$ a new constant scaling solution with negative $u$ is approached from above
\bel{F41}
u=-\frac{1}{128\pi^2}+\frac{3}{16\pi^2}h_\nu^2\rhotil\ .
\ee
We observe that $c_U$ turns negative for $h_\nu^2\rhotil=1/10$ or $k^2=m_\nu^2/5$. Associating roughly $u_0k^4$ with the present dark energy density $\approx(2\cdot10^{-3}\,\text{eV})^4$ yields $k$ in the region of $10^{-2}\,\text{eV}$, rather close to the experimental lower limit for the largest neutrino mass. As $\rhotil$ decreases further $u$ gets more negative due to the large number of fermions in the standard model.

For grand unified theories $c_U$ is typically positive above the unification scale due to the large number of gauge bosons and scalars. With $u_0>0$, negative $u$ between the unification scale and neutrino mass, and positive $u_\infty$ the Einstein potential $\Vhat$ has a rich structure. Approximating $f(\rhotil)=f_0+2\xi_\infty\rhotil$ it approaches a constant for $\rhotil\to0$, $\vp\to-\infty$, decays exponentially for increasing $\vp$ once $2\xi_\infty\rhotil>f_0$, turns negative in the vicinity of the unification scale for $\rhotil\approx10^{4}$, has a minimum for somewhat larger $\vp$ (say $\rhotil\approx10^{5}$) at negative values, turns positive again for $k$ below the neutrino mass ($\rhotil\approx h_\nu^{-2}$) and finally decays exponentially for $\vp\to\infty$. This may lead to interesting features in the cosmological evolution if the solution of the flow equations follows the scaling solution up to $\rhotil\approx10^{116}$, corresponding to $u/f^2)=u/(2\xi_\infty\rhotil)^2\approx10^{-120}$.

\subsection{Scaling solution for curvature coefficient and kinetial}

For a constant $\xi$ we define
\bel{AFS1}
f(\rhotil)=\ftil(\rhotil)+2\xi\rhotil\ .
\ee
Then the scaling solution for $\ftil$ obeys
\bel{AFS2}
\rhotil\partial_{\rhotil}\ftil=\ftil-c_F\ .
\ee
We identify $\xi$ with $\xi_\infty$,
\bel{103A}
\xi=\lim_{\rhotil\to\infty}\frac{f(\rhotil)}{2\rhotil}\ ,\quad \lim_{\rhotil\to\infty}\frac{\ftil(\rhotil)}{2\rhotil}=0\ .
\ee
The scaling solution for $\ftil$ is similar to the one for $u$, with a finite value $\ftil(\rhotil\to\infty)=\ftil_\infty$. Similar to the flow equation for $u$ we approximate
\bel{AFS2A}
c_F=\sum_j\frac{N_j^{(f)}}{96\pi^2(1+\tau_j\rhotil)}+c_F\supt{(grav)}+\frac{N_V^{(0)}}{96\pi^2}\ .
\ee
The numbers $N_j^{(f)}=-N_S^{(j)}-N_F^{(j)}+3N_V^{(j)}$ involve the corresponding numbers of scalars, Majorana fermions and gauge bosons with dimensionless squared mass $\mtil_j^2=\tau_j\rhotil$. For massless gauge bosons the addition of the measure term enhances $2N_V^{(0)}$ to $4N_V^{(0)}$. Again, the Goldstone bosons corresponding to the massive gauge bosons do not contribute.

In this approximation one finds the solution
\bel{AFS2B}
\ftil=\sum_j\ftil_j+\ftil\subt{grav}\ ,
\ee
where
\bel{AFS2C}
\ftil_j=\frac{N_j^{(f)}}{96\pi^2}t_f(\tau_j\rhotil)\ ,
\ee
with threshold function
\bel{AFS2D}
t_f(y)=1+y\ln\left(\frac{y}{1+y}\right)
\ee
obeying
\bel{AFS2E}
y\partial_yt_f=t_f-\frac1{1+y}\ .
\ee

We may write
\bel{AFS3}
\ftil(\rho)=c_F(\rhotil)+\Delta_\rho(\rhotil)\ ,
\ee
where $\Delta_\rho(\rhotil)$ differs from zero only in the threshold regions where the precise $\rhotil$-dependence of $\ftil(\rhotil)$ differs from the one for $c_F(\rhotil)$. For large $\rhotil$ the detailed form of $\ftil$ becomes unimportant, $\ftil(\rhotil)$ being subleading as compared to $2\xi\rhotil$. 

In our approximation the coupling $\xi$ appears only in the metric contribution to the flow equation through $v$ and $\rhotil\partial_{\rhotil}\ln f=1-c_F/(2\xi\rhotil+\ftil)$. For large $\rhotil$ the metric contributions are suppressed by $(\xi\rhotil)^{-1}$. Only the effectively massless particles contribute in this range. The Weyl transformation to the Einstein frame reveals, however, that for a given normalization of $K$, say $K_\infty=\pm1$, the value of $\xi$ enters the wave function renormalization $Z$. The so far neglected mixing of kinetic terms for $\chi$ and the scalar metric fluctuations are important for understanding the precise role of $\xi$ in the region for large $\chi$.

We can also employ eq.~\eqref{F29} for the scaling solution in order to express the wave function renormalization as
\bel{AFS4}
Z=\frac18\left[\frac{K\rhotil}{f}+3\left(1-\frac{c_F}{f}\right)^2\right]\ .
\ee
For large $\rhotil$ we can neglect $c_F/f$ and recover eq.~\eqref{S8}. As long as the flow equation and the scaling form for the kinetial $K(\rhotil)$ is not computed, we can only discuss possible forms which lead to realistic cosmology. The fact that the scaling solution is known for only one of the two scale invariant functions~\eqref{C7} relevant for cosmology, namely $\Vhat(\rhotil)$, clearly limits the predictive power. In principle, the form of $K(\rhotil)$ matters for a precise determination of $\Vhat(\rhotil)$. This effect is small, however, since the relative contribution of the scalar singlet fluctuations to the flow of $U$ and $F$ is small.

\subsection{Robustness of limits of the scaling solution}\label{subsec:LSS}

One may ask how robust are the results for the limiting behavior of the scaling solution for $\rhotil\to0$ and $\rhotil\to\infty$. As we have argued in sect.~\ref{sec:QG} the result~\eqref{Q5},~\eqref{Q7} entails important aspects for the understanding of the overall evolution of the universe. There is actually only a rather limited set of properties that enter this result. First, for any flow equation which ensures a proper decoupling of heavy particles the generic behavior
\bel{F42}
k\partial_k U\sim k^4\ ,\quad k\partial_k F\sim k^2
\ee
is dictated by dimensions. With effective particle masses vanishing (except for threshold regions) the scale $k$ is the only scale relevant for the flow. This assures constant values $u_0$, $f_0$ for $\rhotil\to0$ unless one has a highly non-analytic behavior with $u$ or $f$ increasing $\sim\rhotil^{-1}$ or faster.

Second, there is no good reason why the non-minimal gravitational coupling of the scalar field $\xi$ should be zero for $\rhotil\to\infty$. For positive $\xi_\infty$ one finds $F\sim\xi_\infty\chi^2$ for large $\chi$. Thus for $\rhotil\to\infty$ the fluctuations of the metric become negligible as usually assumed for $k^2<F$. (This assumes a proper diagonalization of the kinetic term in the scalar sector as realized in the Einstein frame.) An exception is a constant contribution to $u$ and $f$ given by the gravitational contribution to $c_U$ and $c_F$. Together with other massless particles this leads to nonzero $c_U$ and $c_F$ for $\rhotil\to\infty$. The constant $c_F$ becomes irrelevant for $f\sim2\xi_\infty\rhotil$.

Third, for the flow of $U$ for $k^2\ll m_\nu^2$ only the effectively massless particles below the neutrino mass scale contribute. These are the photon, the metric fluctuations and the cosmon, unless one extends the standard model to include additional massless particles. The expression for $c_U$ becomes rather simple, with a positive sign for bosons and only counting the number of propagating degrees of freedom. This coincides with simple estimates of the ground state energy in the Hamiltonian formalism. One infers $c_U>0$ for $\rhotil\to\infty$.

Fourth, an increase of $u\sim\rhotil^2$ is not allowed due to stability conditions. This leaves for $u$ only the possibility $u(\rhotil\to\infty)=u_\infty$. A possible exception could only be a value of $v$ very close to one which invalidates the decoupling of the graviton fluctuations for $\partial u/\partial\rhotil$ with $u(\rhotil\to\infty)=4\xi_\infty\rhotil$~\cite{CWIRG}. We will not pursue this possibility here further since it seems not very likely that a full consistent scaling solution can be obtained for this extreme behavior.

Fifth, the metric fluctuations do no longer contribute to $\partial c_U/\partial \rhotil$ of $\partial c_F/\partial\rhotil$ for $\xi_\infty\rhotil\gg1$. With $v=2u/f$,
\bel{F43}
\frac{\partial v}{\partial\rhotil}=\frac{2}{f}\frac{\partial u}{\partial\rhotil}-\frac{2u}{f^2}\frac{\partial f}{\partial\rhotil}=-\frac v\rhotil\ ,
\ee
one obtains expressions which vanish for $v\to0$ as
\bel{F44}
\rhotil\frac{\partial c_U^{(\text{grav})}}{\partial\rhotil}=-v\frac{\partial c_U^{(\text{grav})}}{\partial v}=-\frac{(5+\frac14)v}{128\pi^2}\ .
\ee
In this region only the non-gravitational particles contribute to the field-dependence of $u$ or $f$.

Sixth, the flow of the particle physics couplings all obtain from $\rho$-derivatives of the flow generator for $U$ or $F$. With gravity decoupled the flow of small couplings follows the perturbative $\beta$-functions. Once $\rhotil$ is large enough such that only the particles of the standard model contribute one has a good control of the differential equation defining the scaling solution. The fact that the perturbative $\beta$-functions are reproduced by simply taking derivatives of the flow equation for the effective potential increases the confidence in the validity of this equation.

Taking things together the results for the limiting behavior of the scaling solution seem to be rather robust. The limiting behavior for $\rhotil\to\infty$ only depends on the massless particles. Changes of the constant $c_U(\rhotil\to\infty)$ due to the use of different versions of the flow equations or different forms of the infrared cutoff function can be absorbed by a rescaling of $k$, or a shift of the definition of $\vp$ in the Einstein frame. The fact that $u(\rhotil\to0)$ and $f(\rhotil\to0)$ take constant values seems to be very general. The precise values $u_0$ and $f_0$ typically depend on the unknown particle content in the ultraviolet limit, as well as on the precise implementation of flow equations and cutoff functions. One expects that a large class of models leads to positive $u_0$ and $f_0$.

\section{Crossover cosmology}\label{sec:CC}

In view of the rather robust result for the limiting behavior of the flow equations, as well as for the associated cosmological field equations and their solutions, we investigate in this section in more detail the cosmology describing a crossover between the UV- and IR-fixed points. We focus on the scaling solution according to fundamental scale invariance and discuss possible modifications due to relevant parameters in the next section.

\subsection{Crossover trajectory in scale and time}\label{subsec:CTST}

The scaling solution of quantum gravity exhibits both an ultraviolet (UV) and an infrared (IR) fixed point with the associated quantum scale symmetry. These fixed points are reached for $\rhotil\to0$ and $\rhotil\to\infty$, respectively. For the scaling solution the functions $u(\rhotil)$, $f(\rhotil)$ and $K(\rhotil)$ interpolate between these limits. Such solutions are called \qq{crossover trajectories}. They link two fixed points, describing a crossover from one fixed point behavior to a different one. Along a crossover trajectory the physics can change qualitatively. Close to the fixed points quantum scale symmetry remains a valid approximate symmetry. In regions of a qualitative variation with $\rhotil$ quantum scale symmetry no longer holds since the scale $k$ appears indirectly through $\rhotil$.

One obvious crossover region corresponds to the qualitative change of $f(\rhotil)$ from a constant to a linear increase with $\rhotil$. In the approximation
\bel{CT1}
f(\rhotil)=f_0+2\xi\rhotil\ ,
\ee
the range of this crossover is given by $\rhotil\approx\rhotil_f$, with
\bel{CT2}
\rhotil_f=\frac{f_0}{2\xi}\ .
\ee
The crossover in $f$ describes the onset of the decoupling of the metric fluctuations. For $\rhotil\ll\rhotil_f$ the metric fluctuations play an important role for the flow equations. On the other hand, for $\rhotil\gg\rhotil_f$ the metric fluctuations decouple, their contribution being suppressed by powers of the small quantity $(\xi\rhotil)^{-1}$. (This holds with the exception of a contribution to the $\rhotil$-independent part of the flow of $u$.)

Another possible crossover concerns the behavior of the kinetial $K(\rhotil)$. For the simplified ansatz
\bel{CT3}
K(\rhotil)=\frac\kappa{2\rhotil}+K_0
\ee
the qualitative change occurs for
\bel{CT4}
\rhotil_K=\frac\kappa{2K_0}\ .
\ee
With eqs.~\eqref{CT1} and~\eqref{CT3} one has
\begin{align}
\label{CT5}
Z=&\frac{\chi^2K_E}{16\M}=\frac{K\rhotil}{8f}+\frac3{8f^2}\gl\rhotil\partial_{\rhotil}f\gr^2\nn\\
=&\frac1{16}\left[\frac{\kappa+2K_0\rhotil}{f_0+2\xi\rhotil}+6\left(1+\frac{f_0}{2\xi\rhotil}\right)^{-2}\right]\ ,
\end{align}
which interpolates -- possibly in two steps -- between the limits~\eqref{S25} and~\eqref{S8}. We will see below how a crossover from $Z\gg1$ to $Z\ll1$ could be related to the end of the inflationary epoch in cosmology. Unfortunately, the parameters $\kappa$, $K_0$ and $\xi$ are not yet fixed by the present computations, nor is the approximate form~\eqref{CT3} established.

For the scaling solution of quantum gravity the flow of coupling functions with $k$ translates directly to the dependence of the effective action on the scalar field $\chi$. In turn, the solutions of the field equations derived by variation of this effective action can translate the field dependence into a time dependence. We find typical \qq{crossover cosmologies} for which the infinite past is characterized by $\chi\to0$ and therefore to an approach to the UV-fixed point, while the infinite future realizes $\chi\to\infty$ and therefore approaches the IR-fixed point. The overall simple picture of cosmology is a crossover from the UV-fixed point in the infinite past to the IR-fixed point in the infinite future. The crossover cosmology typically happens in several stages that we will identify with inflation, kination, radiation domination, matter domination and dark energy domination. For the detailed description of the matter and radiation domination epochs particle physics is needed, in our context in the form of the scale invariant standard model and extensions thereof. The inflation and kination epochs can be described by quantum gravity with a scalar field. (Extensions to several scalar fields are possible but will not be discussed in this note.)

\subsection{Inflation and the beginning universe}\label{subsec:IBU}

A rather natural possible beginning of the universe is the close vicinity of the ultraviolet fixed point. Similarly to the infinite increase of $k$ necessary to reach the fixed point precisely, an infinite increase of $-\eta$, which corresponds to an appropriate physical time~\cite{RUCW2}, is needed for the cosmological solution to reach the fixed point. The universe is then eternal, with the fixed point realized precisely only in the infinite past. We discuss here in detail how the vicinity of the UV-fixed point is related to inflation, and how crossover away from the fixed point ends the inflationary epoch.

\zwisch{Beginning at the fixed point}

The UV-fixed point $\chi=0$ is an exact solution of the field equations. This follows generally from the discrete symmetry $\chi\to-\chi$, which only allows even powers of $\chi$ for all terms in the effective action. In the scaling frame for the metric the cosmological evolution equations for $\chi=0$ imply that geometry is given by de Sitter space with conformal Hubble parameter
\bel{IN0}
\cH=-\frac1\eta\ ,\quad \chi=0\ .
\ee
For cosmic time the Hubble parameter $H$ is proportional to $k$,
\bel{IN1}
H^2=\frac{U}{3F}=\frac{u_0k^2}{3f_0}\ .
\ee
Eqs.~\eqref{IN0},~\eqref{IN1} are an exact solution of the field equations for our truncated effective action, provided $\Vhat(\chi=0)>0$. For a more general truncation the expression for $H/k$ may be different, but the partial solution $\chi=0$ remains. Also $\cH=-1/\eta$, which reflects the scale symmetry of de Sitter space, is often realized.

We will see that the solution~\eqref{IN0},~\eqref{IN1} is unstable with respect to small deviations. Arbitrarily small values of $\chi$ will grow as time increases. We describe the beginning stage of the universe by the close vicinity of the solution~\eqref{IN0},~\eqref{IN1}. In the beginning stage only the space averaged field expectation values $g\mnb$ and $\chi$ and the fluctuations of these field encoded in correlation functions matter. Their evolution is very slow if measured in a \qq{physical time} proportional to $\eta$. One may call this stage of the universe~\qq{Great Emptiness}~\cite{CWGE}. In the infinite past $\eta\to-\infty$ the fixed point solution~\eqref{IN1} is approached closer and closer. For $\eta\to-\infty$ all field expectation values vanish since the cosmic scale factor $a(\eta)=(-H\eta)^{-1}$ goes to zero and therefore $g\mnb\to0$. Only the correlation functions differ from zero in this \qq{symmetric} or \qq{ageometric} state.

\zwisch{Inflationary cosmology}

The vicinity of the solution~\eqref{IN0},~\eqref{IN1} corresponds to an epoch of inflationary cosmology, as we will discuss in more detail here. This simple beginning requires $u_0>0$ and $f_0>0$. These conditions are not realized for the standard model of particle physics coupled to quantum gravity for which one finds $u_0<0$, $f_0>0$. We will assume here that the standard model is extended to some grand unified theory at some unification scale much larger than the Fermi scale. Due to the large number of bosonic fields one finds positive $u_0$ for $\text{SO}(10)$- or $\text{SU}(5)$-unification. Any other extension leading to positive $u_0$ is possible as well.

We will discuss inflation in the Einstein frame, since this is most familiar. Many simple features, as the presence of a UV-fixed point and the associated quantum scale symmetry, are no longer directly visible in the Einstein frame. Also the field transformation of the Weyl scaling introduces a mass scale $M$ which is not an intrinsic mass scale for the scaling solution in quantum gravity. On the other hand, the field equations in the Einstein frame take a simple form where one does not need to take the variation of masses into account.

The homogeneous field equations in the Einstein frame take the form
\begin{align}
\label{EF1}
H^2&=\frac1{3\M}\Big( U_E+\frac{Z}2\gl\partial_t\vp\gr^2\Big)\ ,\nn\\
\gl\partial_t^2&+3H\partial_t\gr\vp+\frac{\eta_Z}{8M}\gl\partial_t\vp\gr^2+\frac1Z\frac{\partial U_E}{\partial\vp}=0\ ,
\end{align}
where
\bel{EF2}
\eta_Z=4M\frac{\partial\ln Z}{\partial\vp}=\frac{\partial\ln Z}{\partial\ln\chi}\ .
\ee
The $Z$-factor can be absorbed by using the \qq{canonical scalar field} $\sigma$ with canonical kinetic term, defined by
\bel{EF3}
\frac{\text{d}\sigma}{\text{d}\vp}=Z^{1/2}(\vp)\ .
\ee
For the canonical field the equivalent field equations become
\begin{align}
\label{EF4}
H^2=&\frac1{3\M}\Big( U_E+\frac12\gl\dt\sigma\gr^2\Big)\ ,\nn\\
\gl\dt^2+&3H\dt\gr\sigma=-\frac{\partial U_E}{\partial\sigma}\ .
\end{align}
On the other hand, for a standard form of the potential the physics is encoded in $Z(\vp)$ which often allows for a simple description~\cite{CWCI, GKLR}.

The inflationary epoch is characterized by a slow evolution of the scalar field (\qq{slow roll}) such that the term $\sim\gl\dt\sigma\gr^2$ in eq.~\eqref{EF4} is small as compared to the almost constant $U_E$. Then $H$ is approximately constant such that the expansion becomes exponential or some other very fast increase. For the fixed point solution~\eqref{IN1} the slow roll approximation $\dot{H}/H^2\ll1$, $\gl\dt\sigma\gr^2\ll U_E$ becomes exact. In the Einstein frame this corresponds to $\vp\to-\infty$ for which the potential approaches a constant value
\bel{EF5}
U_E\to\frac{u_0M^4}{f_0^2}\ .
\ee
In the vicinity of the UV-fixed point the slow roll approximation remains valid.

The slow roll approximation is characterized by two small parameters
\begin{align}
\label{EF6}
\eps=&\frac{\M}{2}\left(\frac{\partial\ln U_E}{\partial\sigma}\right)^2\ ,\nn\\
\eta=&\frac{\M}{U_E}\frac{\partial^2U_E}{\partial\sigma^2}=2\eps+\M\frac{\partial^2\ln U_E}{\partial\sigma^2}\ .
\end{align}
Inflation ends once $\eps$ or $\eta$ reach values of the order one. In terms of $\vp$ or $\rhotil$ the slow roll parameters are given by
\begin{align}
\label{EF7}
\eps=&\frac1{2Z}\left(M\frac{\partial\ln U_E}{\partial\vp}\right)^2=\frac1{8Z}\left[\rhotil\partial_{\rhotil}\ln\left(\frac{u}{f^2}\right)\right]^2\ ,\nn\\
=&\frac1{2Z}\left(\frac{c_F}{f}-\frac{c_U}{u}\right)^2\ ,
\end{align}
and
\begin{align}
\label{EF7A}
\eta=&2\eps+\frac{\M}{Z}\sder{\ln U_E}{\vp}-\frac{\eta_Z}{8Z}M\der{\ln U_E}{\vp}\nn\\
=&2\eps+\frac1{4Z}\left[\rhotil^2\partial_{\rhotil}^2+\gl1-\frac{\eta_Z}{4}\gr\rhotil\partial_{\rhotil}\right]\ln\left(\frac{u}{f^2}\right)\nn\\
=&\frac1Z\bigg\{\gl1+\frac{\eta_Z}{8}\gr\frac{c_U}{u}-\frac12\gl1+\frac{\eta_Z}{4}\gr\frac{c_F}{f}\nn\\
&+\frac{3c_F^2}{2f^2}-\frac{2c_Fc_U}{fu}+\frac{1}{2f}\rhotil\partial_{\rhotil}c_F-\frac{1}{2u}\rhotil\partial_{\rhotil}c_U\bigg\}\ .
\end{align}
For the last relations for $\eps$ and $\eta$ we have employed the scaling equation~\eqref{F29}. This demonstrates directly that the slow roll parameters are calculable in terms of the scaling solution of the flow equations!

For $\vp\to-\infty$ or $\rhotil\to0$ we may use the limit of the scaling solution
\bel{EF8}
u=u_0+\mtil_0^2\rhotil\ ,\quad f=f_0+2\xi_0\rhotil\ ,
\ee
or
\begin{align}
\label{EF9}
U_E&=\frac{M^4u}{f^2}=\frac{M^4u_0}{f_0^2}\left[1+\beta_0\rhotil\right]\ ,\nn\\
\beta_0&=\frac{\mtil_0^2}{u_0}-\frac{4\xi_0}{f_0^2}\ .
\end{align}
With
\begin{equation}
\label{EF10}
\rhotil=\frac{\chi^2}{2k^2}=\frac12\exp\left(\frac{\vp}{2M}\right)\ ,\quad M\der{\ln U_E}{\vp}=\frac{\rhotil}{2}\der{\ln U_E}{\rhotil}\ ,
\end{equation}
one finds for $\rhotil\to0$ or $\vp\to-\infty$
\bel{EF10A}
M\der{\ln U_E}{\vp}=\frac{\beta_0\rhotil}{2}=\frac{\beta_0}{4}\exp\left(\frac{\vp}{2M}\right)\ .
\ee
Thus $\eps$ and $\eta$ vanish exponentially for $\vp\to-\infty$. We conclude that the scaling solution predicts an inflationary epoch in cosmology.

\zwisch{End of inflation by kinetial crossover}

We have argued in sect.~\ref{sec:QSS} that the observed small size of the primordial cosmic fluctuations requires that fluctuations decouple when $\vp$ is already larger than $M$, unless the ratio $u_0/f_0^2$ is tiny. In this case the potential is already exponentially decreasing according to eq.~\eqref{S7A}, providing for a natural suppression factor for the fluctuations. In the range of validity of eq.~\eqref{S7A} one has $M\partial\ln U_E/\partial\vp=-1$, such that $\eps$ and $\eta$ only depend on $Z$~\cite{CWIQM}
\bel{EF11}
\eps=\frac1{2Z}\ ,\quad \eta=\frac{1+8\eta_Z}{Z}\ .
\ee
These simple relations make a discussion of inflationary cosmology in terms of the kinetial $Z(\vp)$ very convenient.

In the approximation~\eqref{EF11} an inflationary epoch lasts as long as $Z$ remains larger than one. If the end of inflation occurs for a range of $\vp$ for which the exponential decrease~\eqref{S7A} is valid there is a possibility to associate the end of inflation with a crossover in the kinetial~\eqref{CT5}
\bel{EF12}
Z=\frac38+\frac{K_0}{16\xi}+\frac{\kappa}{32\xi\rhotil}\ .
\ee
Values of $Z$ smaller than one can happen for large enough $\rhotil$, in particular for negative $K_0$ or, more generally, if $K(\rhotil)$ reaches negative values. It is conceivable that a crossover in $Z$ happens for values of $\rhotil$ much larger than one. This could explain a small value
\bel{EF13}
\frac{U_E}{M^4}=\frac{u}{f^2}=\frac{u}{4\xi^2\rhotil^2}\ ,
\ee
and therefore a small amplitude of the primordial fluctuations~\eqref{46A}.

\zwisch{End of inflation by grand unified threshold}

The approximation~\eqref{EF11} is valid if $u(\rhotil)$ and $f(\rhotil)/\rhotil$ are approximately constant. This may not hold in threshold regions where some of the particles decouple due to their $\rhotil$-dependent mass. We will next argue that a threshold region is a good candidate for ending inflation. In view of the small observed fluctuation amplitude we can use $u/f^2\ll1$ and $|c_F/f|\ll1$ in order to simplify the general expressions~\eqref{EF7},~\eqref{EF7A} for the slow roll parameters
\begin{align}
\label{EF13A}
\eps&=\frac{c_U^2}{2Zu^2}\ ,\nn\\
\eta&=\frac1Z\Big\{\gl1+\frac{\eta_Z}{8}\gr\frac{c_U}{u}-\frac1{2u}\rhotil\partial_{\rhotil}c_U\Big\}\ .
\end{align}
In the flat regions for $u$ one has $u=c_U=\text{const.}$ and we recover eq.~\eqref{EF11}. In the threshold regions $c_U/u$ can differ from one substantially, however. For $|c_F/f|\ll1$ one has
\begin{equation}
\label{EF13B}
c_U=\frac{\Nbar_U(\rhotil)}{128\pi^2}\ ,\quad \Nbar_U=2+\Nbar_S+2\Nbar_V-2\Nbar_F\ ,
\end{equation}
where the effective particle numbers can change rather rapidly with $\rhotil$.

As an example we consider the variation of $c_U$ for the transition from some grand unified theory (GUT) to the standard model (SM) at a scale
\bel{EF13C}
m_X=\delta\sqrt{F}\ ,\quad \mtil_X^2=\frac{m_X^2}{k^2}=2\delta^2\xi\rhotil\ .
\ee
In the Einstein frame the small parameter $\delta$ denotes the ratio of the GUT-scale $M_X$ over the Planck mass $M$. For the standard model + cosmon one has $\Nbar_S=5$, $\Nbar_V=12$, $\Nbar_F=45$, and therefore negative $\Nbar_U$,
\bel{EF13D}
\Nbar_U^{\text{(SM)}}=2+5+24-90=-59\ ,
\ee
while $\text{SO}(10)$-unification implies $\Nbar_V=45$, $\Nbar_F=48$, leading to positive $\Nbar_U$,
\bel{EF13E}
\Nbar_U^{\text{(GUT)}}=\Nbar_S^{\text{(GUT)}}-6\ .
\ee
The number of scalars $\Nbar_S^{\text{(GUT)}}$ depends on the particular $\text{SO}(10)$-model and is typically large.

For a simplified model of the threshold all non-SM particles are taken to have the same mass, resulting in
\bel{EF13F}
\Nbar_U(\rhotil)=\frac{\Nbar_S\supt{(GUT)}+53}{1+\mtil_X^2}-59\ ,
\ee
or
\bel{EF13G}
c_U=\frac1{128\pi^2}\left(\frac{A}{1+\tau\rhotil}+B\right)\ ,
\ee
with
\bel{EF13H}
A=\Nbar_S\supt{(GUT)}+53\ ,\quad B=-59\ ,\quad \tau=2\delta^2\xi\ .
\ee
Away from the threshold region where $\tau\rhotil\approx1$ one expects flat regions, with $u=(A+B)/(128\pi^2)$ for $\tau\rhotil\ll1$ and $u=B/(128\pi^2)$ for $\tau\rhotil\gg1$. Since $u$ changes from positive to negative values as $\rhotil$ increases there will be a region where $u\approx0$ and therefore $\eps\gg1$. Such a threshold will end the inflationary epoch.

\zwisch{Threshold behavior}

For a more detailed picture we need to compute the $\rhotil$-dependence of $u$ through the threshold region~\eqref{88C},~\eqref{88D},
\bel{EF13GA}
u=\frac1{128\pi^2}\left(At_u\gl\tau\rhotil\gr+B\right)\ .
\ee
It interpolates smoothly between
\bel{EF13HA}
u\gl\tau\rhotil\ll1\gr=\frac{A+B}{128\pi^2}\ ,\quad u\gl\tau\rhotil\gg1\gr=\frac{B}{128\pi^2}\ .
\ee
The parameter $\tau$ sets only the position of the threshold, since it can be absorbed by a shift in $x=\ln\rhotil$. For $B<0<A+B$ $u$ changes sign at $\bar x$ where $c_U$ is still positive. Indeed, with $c_U-u=-\frac12\rhotil\partial_{\rhotil}u$ the r.h.s. is positive if $u(\rhotil)$ is decreasing. Starting from small $\rhotil$ both $u$ and $c_U$ decrease while $c_U/u$ increases for increasing $\rhotil$
\bel{EF13J}
\frac{c_U}{u}=1+\frac{A\tau \rhotil}{A+B}\ .
\ee
The increase of $c_U/u$ continues until it diverges as $u\to0$ for $\rhotil\to\bar \rho$. The slow roll approximation breaks down for $\rhotil<\bar \rho$. For $\rhotil>\bar \rho$ the potential $u$ becomes negative and $\tilde u$ approaches the negative constant $B$ for $\rhotil\to\infty$. The coefficient $c_U$ changes signs for $\rhotil>\bar \rho$.

\begin{figure}[h]
\includegraphics{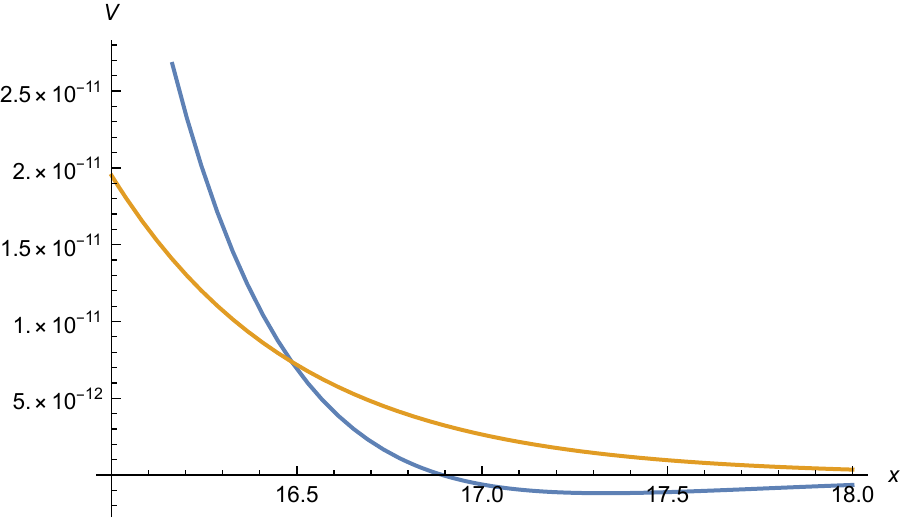}
\caption{\emph{Minimum of scalar potential for grand unified theory}. We show $\Vhat$ as a function of $x=\ln\rhotil=\vp/(2M)$ in a range of large $x$. One observes the shallow minimum for the blue curve. For comparison, we also plot the potential that would be obtained by moving the Fermi scale to the grand unified scale (orange curve, for $\Vhat/10$ for better visibility).}
\label{fig:2}
\end{figure}

As a consequence, the frame invariant potential $\Vhat$ or the potential in the Einstein frame $U_E=M^4\Vhat$ exhibits a shallow minimum for $\rhotil>\bar\rho$. We show this in fig.~\ref{fig:2}
for a typical grand unified theory based on $\text{SO}(10)$. Realistic cosmology has to avoid that the scalar field settles in this minimum after inflation. Otherwise, a substantial negative cosmological constant would stop further expansion.

\zwisch{Grand unified scale}

The scaling solution predicts that inflation ends at the latest during the GUT-phase transition when $u$ turns negative due to the large number imbalance between fermions and bosons in the standard model. This relates naturally the small fluctuation amplitude to the small ratio $\delta$ between the unification mass $M_X$ and the Planck mass $M$. Indeed, the characteristic range for this threshold and therefore the end of inflation is given by $\rhotil_e\approx\tau^{-1}=\gl2\delta^2\xi\gr^{-1}$. At this value of $\rhotil_e$ one has
\bel{EF13K}
\Vhat_e=\frac{u_e}{4\xi^2\rhotil_e^2}\approx\delta^4u_e\ ,
\ee
with $u_e=u(\rhotil_e)$ and $\xi$ characterizing $f(\rhotil_e)=2\xi\rhotil_e$. Typical values $M_X\approx10^{16}\,\text{GeV}$, with $\delta\approx10^{-2}$, are well compatible with grand unified models and lead to a realistic amplitude of the primordial cosmic fluctuations.

\subsection{Primordial fluctuation spectrum}\label{subsec:PF}

The quantum effective action determines the full propagator or the connected two-point correlation function as the inverse of its second functional derivative. For a Minkowski signature taking the inverse amounts to an initial value problem~\cite{CWCFEA}. One may assume that the propagator takes for very large (comoving and covariant) momenta the same Lorentz-covariant form as for flat space. This is well motivated since effects of a non-flat geometry become suppressed for wave lengths of the fluctuations much smaller than characteristic geometric length scales. With this assumption the propagators for the physical metric fluctuations and scalar fluctuations are the same as usually obtained by canonical quantization in a Bunch-Davies vacuum~\cite{BUDA}. The amplitude and spectrum of the primordial cosmic fluctuations are directly given by the full propagator. Computing the quantum effective action therefore gives direct quantitative access to the cosmic fluctuation spectrum~\cite{CWCFEA, CWMF, CWCFVG}. We will assume here the approximation~\eqref{C3} with $F$, $U$ and $K$ according to the scaling solution for quantum gravity.

The cosmic fluctuation spectrum and the amplitude of the fluctuations are independent of the metric frame~\cite{CWCFVG}. We can therefore turn immediately to the well known results for the inflationary epoch in the Einstein frame. The amplitude of the scalar fluctuations $\cA$ and tensor fluctuations $r\cA$ is given by eqs.~\eqref{S27}-~\eqref{46A}, where $\cA$ can be compared with observation. For inflationary cosmology the tensor to scalar ratio $r$ is given by
\bel{EF14}
r=16\eps\ .
\ee
Furthermore, the momentum dependence of the scalar fluctuations is determined by a spectral index
\bel{EF15}
n=1-6\eps+2\eta\ .
\ee
Once $Z(\vp)$ is computed for the scaling solution both $r$ and $n$ are determined for a given particle content. Observation of the primordial fluctuation spectrum can be used directly for a test of models.

The quantities $U_E/M^4$, $\eps$ and $\eta$ relevant for the observed fluctuations of the CMB have to be evaluated at a time when their wave lengths decouple at a certain number $N$ of $e$-foldings before the end of inflation (typically $N\approx 50-70$ depending on the heating after inflation). In the Einstein frame this decoupling happens when the corresponding wave length exceeds the horizon. The same decoupling happens in all metric frames even though no geometric horizon may be present. Computing $\vp(N)$ at this time expresses $r$ and $n$ as functions of $N$. For this purpose one replaces the time variable $t$ by the number of $e$-foldings, with $a_f$ the scale factor at the end of inflation
\bel{138A}
N=-\ln\left(\frac{a}{a_f}\right)\ ,\quad \frac{\text{d}N}{\text{d}t}=-H\ .
\ee
The field equation in the slow roll approximation reads
\bel{138B}
\frac{\partial}{\partial N}\left(\frac\sigma M\right)=M\der{\ln U_E}{\sigma}\ ,
\ee
and one infers the expression (with $\vp_f$ the value of $\vp$ at the end of inflation)
\begin{align}
\label{138C}
N=&\frac1M\int_\vp^{\vp_f}\text{d}\vp'Z(\vp')\left(-M\der{\ln U_E}{\vp'}\right)^{-1}\nn\\
\approx&\frac1M\int_\vp^{\vp_f}\text{d}\vp'Z(\vp')\frac{u}{c_U}(\vp')\ .
\end{align}
This expression simplifies~\cite{CWIQM} for flat regions in $u$ where $u=c_U$.

In the approximation~\eqref{EF11} one has $\eps=8/Z$ and $n=1-(1-16\eta_Z)/Z$. This requires a large value of $Z$ at the time of decoupling of the fluctuations. Realistic values of $r/(1-n)$ also demand substantial negative $\eta_Z$. The relations~\eqref{EF11} will be modified according to eq.~\eqref{EF13A}, if decoupling occurs in a threshold region. Given the high predictivity of the scaling solution for quantum gravity it is possible that the simplest model of the metric coupled to a single scalar field may be falsified by observation once a computation of the kinetial becomes available. One may then have to proceed to extensions with more than one scalar field or invariants with more than two derivatives playing a role during inflation.

\subsection{Kination}\label{subsec:K}

The inflationary epoch is typically followed by an epoch for which the scalar kinetic energy dominates, while radiation and matter are still negligible. 

\zwisch{Scaling solution for kination}

The kination epoch can be characterized by a scaling solution~\cite{Wetterich_1988}. For this solution the field $\sigma$ changes logarithmically with time, such that the kinetic energy decreases $\sim t^{-1}$. We make the ansatz
\bel{K1}
\sigma=cM\ln\left(\frac{t}{t\subt{kin}}\right)\ ,\quad H=\eta_Ht^{-1}\ .
\ee
For this ansatz the field equations take the form
\begin{align}
\label{K2}
\eta_H^2&=\frac{c^2}{6}+\frac{U_Et^2}{3\M}\ ,\nn\\
3\eta_H-1&=-\frac{t^2}{cM}\der{U_E}{\sigma}\ .
\end{align}
As long as the potential $U_E$ can be neglected the solution for the kination period reads
\bel{K3}
\eta_H=\frac13\ ,\quad c^2=\frac23\ .
\ee

The kination approximation~\eqref{K1}-~\eqref{K3} remains valid as long as the contribution from the effective potential or from radiation and non-relativistic matter remains negligible. In the limit of constant $Z$ one has $\vp=Z^{-1/2}\sigma$ and therefore
\begin{align}
\label{K4}
U_E\sim &M^4\exp\left(-\frac{\vp}{M}\right)\nn\\
=&M^4\exp\left(-\frac{\sigma}{\sqrt{Z}M}\right)\sim\left(\frac{t}{t\subt{kin}}\right)^{-\sqrt{\frac{2}{3Z}}}\ .
\end{align}
Thus $U_E$ decays faster than $t^{-2}$ if $Z<1/6$. If such a regime with constant $Z<1/6$ is reached the potential becomes less and less important as $t$ increases. This also holds for
\bel{K4A}
M\der{U_E}{\sigma}=-\frac{U_E}{\sqrt{Z}}\ .
\ee

On the other hand, in the approximation of constant $\eta_Z=4M\partial\ln Z/\partial\vp$ one has
\begin{align}
\label{K5}
Z&=\Zbar \exp\left(\frac{\eta_Z\vp}{4M}\right)\ ,\nn\\
\sigma&=\frac{8M\Zbar^{1/2}}{\eta_Z}\exp\left(\frac{\eta_Z\vp}{8M}\right)+\sigma_0\ .
\end{align}
For $\eta_Z<0$ the integration constant $\sigma_0$ is the value of $\sigma$ that is reached for $\vp\to\infty$. With
\bel{K6}
U_E\sim\exp\left(-\frac{\vp}{M}\right)=\left[\frac{\eta_Z}{8M\sqrt{\Zbar}}(\sigma-\sigma_0)\right]^{-\frac{8}{\eta_Z}}\ ,
\ee
\bel{K7}
M\der{U_E}{\sigma}=-\frac{U_E}{\sqrt{Z}}\sim\Zbar^{-\frac12}\left[\frac{\eta_Z}{8M\sqrt{\Zbar}}(\sigma-\sigma_0)\right]^{-\frac{8}{\eta_Z}-1}\ ,
\ee
and logarithmically increasing $\sigma/M=\sqrt{2/3}\ln(t/t\subt{kin})$ one finds for $\eta_Z>0$ that the contributions $\sim U_Et^2$ in eq.~\eqref{K2} can no longer be neglected for large enough $t$. For $\eta_Z<0$ the field $\sigma$ would reach $\sigma_0$, and therefore $\vp$ diverges, at a finite time $t\subt{end}$.

A numerical solution of the cosmological field equations~\eqref{EF1} shows that for $Z$ of the order one the kination epoch is short and the scalar field does not grow to values far beyond the minimum of $U$. In contrast, for a crossover in $Z$ to values smaller than $1/6$ the kination epoch will stop only due to the presence of radiation and matter. In the absence of radiation and matter it is a cosmic attractor solution. The scalar potential $\Vhat$ can decrease to very small values during this epoch.

\zwisch{Transition to radiation and matter domination}

Assume that after inflation a certain amount of relativistic particles is created. This produces entropy and heats the universe, with particles forming locally an equilibrium state with temperature $T$. (For a discussion of heating in similar models see ref.~\cite{RUCW}.) The radiation density $\rho_r$ decreases $\sim a^{-4}$, in contrast to the faster decrease of the scalar kinetic energy density $\gl\dt\sigma\gr^2\sim t^{-2}\sim a^{-6}$. The kination epoch ends once the radiation energy equals the scalar kinetic energy. If in the following radiation dominated epoch the scalar potential can be neglected, one has
\bel{RD1}
H=\frac1{2t}\ ,\quad \dt^2\sigma+\frac3{2t}\dt\sigma=0\ ,
\ee
with solution
\bel{RD2}
\dt\sigma\sim t^{-3/2}\ ,\quad \gl\dt\sigma\gr^2\sim t^{-3}\sim a^{-6}\ .
\ee
Thus the ratio $\gl\dt\sigma\gr^2/\rho_r\sim a^{-2}$ continues to decrease.

During radiation domination the scalar field evolves only slowly
\bel{RD3}
\sigma(t)=\sigma_r+2c_rt_r\left(1-\sqrt{\frac{t_r}{t}}\right)\approx\sqrt{\frac83}M\left(1-\sqrt{\frac{t_r}{t}}\right)\ ,
\ee
with
\bel{RD4}
c_r=\gl\dt\sigma\gr(t_r)\approx\sqrt{\frac23}\frac{M}{t_r}\ ,\quad \sigma_r=\sigma(t_r)\ ,
\ee
the initial conditions at the onset of radiation domination. The evolution of the scalar field almost stops, approaching $\bar\sigma=\sigma_r+2c_rt_r$. Correspondingly, the scalar potential undergoes only a small change
\bel{RD5}
U_E=U_E(t_r)\exp\bigg[-\sqrt{\frac{8}{3Z}}\left(1-\sqrt{\frac{t_r}{t}}\right)\bigg]\ .
\ee

The overall picture is simple. The presence of radiation essentially stops the further evolution of the scalar field which settles at the value it has reached at the onset of radiation domination. For a realistic cosmology radiation domination has to set in before nucleosynthesis. Otherwise the different time history due to a substantial kinetic energy of the scalar field would modify the element abundances. For an end of the kination epoch close to nucleosynthesis small changes of abundances are expected.

Due to a baryon asymmetry that has to be created at some moment radiation domination will be replaced by matter domination. The evolution of the scalar field during matter domination is qualitatively similar to radiation domination. The overall picture is that the evolution of the scalar field becomes very slow such that the kinetic energy $T_E=\gl\dt\sigma\gr^2/2$ becomes comparable to $U_E$ at some time $t_S$. The evolution of the scalar field beyond $t_S$ depends on the value of $U_S=U_E(t_S)$. If $\sigma(t_S)$ is in a range where $U_S$ is negative the gradient term $\sim\partial U_E/\partial\sigma$ in the field equation leads to a subsequent decrease of the scalar field, with $U_E$ becoming more negative as $t$ increases beyond $t_S$. Cosmology with negative $U_E$ in the present epoch is not comparable with observation. In contrast, for positive $U_S$ the energy density of the scalar field constitutes a form of dynamical dark energy.

\subsection{Dynamical dark energy}\label{subsec:DDE}

Due to the role of the neutrino fluctuations for the scaling solution of $u(\rhotil)$ the potential $U_E(\vp)$ has a local maximum at a positive values, $U\subt{max}=U_E(\vp\subt{max})>0$. For $U_S>U\subt{max}$ the scalar field continues to increase and will reach asymptotically a (approximate) scaling solution~\cite{Wetterich_1988} for (approximately) constant $Z$. For $U_S<U\subt{max}$ the scalar field typically reaches a turning point and decreases afterwards. Only for $U_S$ sufficiently close to $U\subt{max}$ it may reach $U\subt{max}$ and turn towards the scaling solution. The precise dynamics of dark energy depends on the ratio $U_S/U\subt{max}$ and on $Z$.

\zwisch{Cosmological scaling solution}

For constant $Z$ the scalar field equation reads ($\sigma=\sqrt{Z}\vp$, $\eta_Z=0$)
\bel{DS1}
\gl\dt^2+3H\dt\gr\sigma=-\der{U_E}{\sigma}=\frac{uM^3}{\sqrt{Z}\xi^2}\exp\left(-\frac{\sigma}{\sqrt{Z}M}\right)\ ,
\ee
where
\bel{DS2}
U_E=\frac{uM^4}{\xi^2}\exp\left(-\frac{\sigma}{\sqrt{Z}M}\right)\ .
\ee
The Hubble parameter obeys
\bel{DS3}
H^2=\frac1{3\M}\big[\rho+U_E+\frac12\gl\dt\sigma\gr^2\big]\ ,
\ee
with
\bel{DS4}
\dt\rho+nH\rho=0\ .
\ee

For cosmological scaling solutions~\cite{Wetterich_1988, CWCMAV, CLW} the dark energy density follows the same time dependence as the dominant radiation ($n=4$) or matter ($n=3$) energy density, with
\bel{DS5}
\rho=\frac{\rho_0\M}{t^2}\ ,\quad H=\frac2{nt}\ .
\ee
The scalar field changes logarithmically
\bel{DS6}
\sigma=\sigma_0+c_\sigma M\ln\left(\frac{t}{t_0}\right)\ ,
\ee
such that eq.~\eqref{DS1} becomes
\bel{DS7}
\gl\frac6n-1\gr c_\sigma t^{-2}=\frac{u\M}{\sqrt{Z}\xi^2}\exp\left(-\frac{\sigma_0}{\sqrt{Z}M}\right)\left(\frac{t}{t_0}\right)^{-\frac{c_\sigma}{\sqrt{Z}}}\ .
\ee
For positive $u$ one has $c_\sigma>0$, such that eq.~\eqref{DS7} is obeyed for
\bel{DS8}
c_\sigma=2\sqrt{Z}\ ,
\ee
and
\bel{DS9}
\frac{u\M t_0^2}{\xi^2}\exp\left(-\frac{\sigma_0}{\sqrt{Z}M}\right)=2\gl\frac6n-1\gr Z\ .
\ee
From eq.~\eqref{DS3} one infers
\bel{DS10}
\rho_0=\frac{12}{n}\gl\frac1n-Z\gr\ .
\ee
The fraction of homogeneous dark energy is given by
\bel{DS11}
\Omega_h=\frac{U_E+\frac12\gl\dt\sigma\gr^2}{3\M H^2}=Zn\ .
\ee
The scaling solution is a cosmic attractor solution in the sense that neighboring solutions approach it for increasing $t$~\cite{Wetterich_1988, CWCMAV}. For negative $u$ no scaling solution of this type exists.

\zwisch{Quintessence}

Quantum gravity predictions for dynamical dark energy or quintessence are only partial. The scaling solution predicts the potential $u(\rhotil)$ rather accurately. On the other hand, the translation to $U_E(\sigma)$ involves the curvature coefficient $f(\rhotil)$ and the wave function renormalization $Z(\rhotil)$. Even in the approximation where we only employ two constants $\xi_\infty$ and $Z$, with $f(\rhotil\gg1)=2\xi_\infty\rhotil$, the dynamics of quintessence further depends on $U_S$ or, equivalently, $\sigma_S$. We may interpret these values in terms of \qq{initial conditions} for the evolution in the present epoch. In principle, $\sigma_S$ is computable if $u(\rhotil)$, $f(\rhotil)$ and $K(\rhotil)$ are known. The duration of the kination epoch depends on the detailed dynamics near the end of inflation. For given $u$, $f$, $K$ inflation is an attractor solution which has no free parameters. All quantities are determined as functions of the number of $e$-foldings before the end of inflation on which the dynamics depends. In practice, however, $\sigma_S$ depends on too many details. We may take the inverse attitude and consider $\sigma_S/M$ as a free dimensionless parameter. If a value of $\sigma_S$ leads to realistic cosmology we may use this as a constraint on the detailed physics near the end of inflation. In this sense $\sigma_S$ \qq{monitors} the quantum fluctuations at scales relevant for the end of inflation. 

The value of the potential at the maximum $U\subt{max}$ is determined for $f=2\xi\rhotil$ by the zero of $c_U$,
\bel{AQ1}
\rhotil\der{\Vhat}{\rhotil}=\frac1{f^2}\gl\rhotil\partial_{\rhotil}u-2\gr=-\frac{2c_U}{f^2}=0\ .
\ee
According to eq.~\eqref{F39} this is given by ($h\nb^2=\xi m\nb^2/M^2$)
\bel{AQ2}
2h\nb^2\rhotil\subt{max}=\frac15\ ,\quad \rhotil\subt{max}=\frac{M^2}{10\xi m\nb^2}\ ,
\ee
with
\bel{AQ3}
M^4\exp\left(-\frac{\vp\subt{max}}{M}\right)=25\xi^2m\nb^4\ .
\ee
The maximum value of the potential is given by the neutrino mass
\bel{AQ4}
U\subt{max}=25u(\rhotil\subt{max})m\nb^4\ ,
\ee
where
\bel{AQ5}
u(\rhotil\subt{max})=\frac1{128\pi^2}\gl5-6t_u(0.2)\gr=\frac{0.54}{128\pi^2}=4.27\cdot10^{-4}\ ,
\ee
such that
\bel{AQ6}
\frac{U\subt{max}^{1/4}}{m\nb}=0.32\ .
\ee
It is striking that the characteristic scale for dynamical dark energy turns out very close to the neutrino mass, with details depending on the precise mass pattern for the three neutrinos. On the other hand, this makes the detailed understanding of the quintessence dynamics more involved. We observe that a constant value $\sigma(t)=\sigma\subt{max}$ is an exact solution, with effective cosmological constant given by $U\subt{max}$ not very far from the observed dark energy density. For fundamental scale invariance the solution $\sigma=\sigma\subt{max}$ is the only solution for which dark energy is a constant. The generic prediction is a dynamical form of dark energy rather than an effective cosmological constant. For cosmologies in the vicinity of this solution one may expect a slow evolution. Further complexity may arise if a crossover in the neutrino sector induces in the Einstein frame a $\sigma$-dependence of the neutrino masses, as for growing neutrino quintessence~\cite{AQCGM, CWGNCS, ABFPW, MPRC, CPW}. Within the scaling solution this happens if $h\nb$ depends on $\rhotil$. In this case a possible cosmic scaling solution ends once the neutrinos become non-relativistic.

\zwisch{Early dark energy}

Fundamental scale invariance has an important consequence for the possible existence of early dark energy. For the epochs of radiation and matter domination the scalar potential is bounded by $U\subt{max}$, typically with $U_S$ having a similar order of magnitude as $U\subt{max}$. For the epochs of radiation-matter equality or last scattering the energy density $\rho_E$ in radiation and matter obeys $\rho_E\leq0.2\,\text{eV}$. This imposes for these epochs a bound on the fraction of early dark energy $\Omega\subt{EDE}$,
\bel{EDE1}
\Omega\subt{EDE}\leq\frac{U\subt{max}}{\rho_E}\approx6.5\left(\frac{m\nb}{\text{eV}}\right)^4\ .
\ee
During these epochs the kinetic energy of the scalar field is already tiny. According to eq.~\eqref{RD2} its relative fraction has decreased by a factor $\sim\gl a\subt{NS}/a\gr^2$ since nucleosynthesis.

\section{Fundamental scale invariance}\label{sec:FSI}

Fundamental scale invariance~\cite{CWFSI} states that the world is described by the exact scaling solution of functional flow equations for quantum gravity. This property follows for a theory without any intrinsic length or mass scale. A theory with fundamental scale invariance can be formulated entirely in terms of \qq{scale invariant fields} without any appearance of $k$. This includes the existence of a continuum limit. In consequence, the quantum effective action does not involve $k$ once expressed in terms of the scale invariant fields. In our context the scale invariant fields are $\tilde\chi=\chi/k$ and $\gtil\mnb=k^2g\mnb$. Indeed, the use of $\gtil\mnb$ absorbs the factors $k^4$ in $U$ and $k^2$ in $F$ in eq.~\eqref{C3}. Expressing the scaling solution in terms of scale invariant fields the effective action no longer involves $k$,
\bel{FS1}
\Gamma=\int_x\sqrt{\gtil}\bigg\{-\frac{f(\rhotil)}{2}\tilde R+\frac12K(\rhotil)\partial\mb\tilde\chi\partial\nb\tilde\chi\gtil\mno+u(\rhotil)\bigg\}\ .
\ee

For computing the scaling functions $u(\rhotil)$, $f(\rhotil)$ or $K(\rhotil)$ for a theory with fundamental scale invariance we have employed functional flow equations for the variation of an effective infrared cutoff $R_k$. This concept can be employed for a formulation in terms of scale invariant fields as well~\cite{CWFSI}. The flow equation~\eqref{F1} formulates the $k$-dependence of an effective action for which only fluctuations with squared momenta $q^2>k^2$ are included. It is formulated for fixed fields $\chi$, $g\mnb$ such that the infrared cutoff requires for the momenta of fluctuations that are effectively included in $\Gamma_k$ $\bar q^2=q\mb q\nb g\mno/\chi^2\gtrsim k^2/\chi^2$. (Roughly speaking the dimensionless quantity $\bar q^2$ is the squared momentum in units of the $\chi$-dependent masses.) In terms of the scale invariant fields this condition becomes $\bar q^2\gtrsim\tilde\chi^{-2}$. No scale $k$ appears anymore, while the fluctuation effects are no studied for varying field values $\tilde\chi$. The flow equation describes how the effective action changes as additional fluctuations are included due to changing dimensionless masses $\sim\tilde\chi$. This is the content of the differential equations~\eqref{F29} for the $\rhotil$-dependence of the scaling solutions.

Fundamental scale invariance requires the existence of a scaling solution. In the other direction, the existence of a scaling solution guarantees the existence of an effective action which is compatible with fundamental scale invariance. We have already observed that for the scaling solution the scale $k$ is no longer present if we formulate the field equations in the Einstein frame. In view of the scale invariant formulation~\eqref{FS1} this should not be a surprise.

\subsection{Predictivity}\label{subsec:P}

Theories with fundamental scale invariance have a high predictive power. As compared to general renormalizable quantum field theories formulated in terms of a UV-fixed point the relevant parameters for the flow away from the scaling solution are absent. This yields important additional restrictions in the space of all possible renormalizable quantum field theories. Free parameters arise only if there exists a whole family of scaling solutions that can be parameterized by these parameters.

In practical terms the restrictions arise because the scaling solutions have to exist for the whole range of fields, momenta etc. Concerning our truncation the scaling solutions for $u$, $f$ and $K$ have to exist for the whole range of $\rhotil$ from zero to infinity. Properties at both ends of the interval $0\leq\rhotil<\infty$ matter for the existence of solutions, somewhat similar to the possible solutions of the Schrödinger equation for radial wave functions in the hydrogen atom. For the region of small $\rhotil$ we may consider potential scaling solutions of eq.~\eqref{F29} as an initial value problem, with initial data set at $\rhotil=0$. Not all of the initial data lead to solutions that can be extended to the whole range of $\rhotil$. Furthermore, we know that for $\rhotil\to\infty$ the function $u(\rhotil)$ has to approach a constant $u_\infty$, and $f(\rhotil)/\rhotil$ has to reach a constant value $2\xi_\infty$.

These conditions put severe restrictions on the scaling solutions for models with a given content of particles. This holds, in particular, for the range of large $\rhotil$ for which the particle content is restricted by observation of the \qq{low energy physics}. For the example of the standard model, the scaling solution puts an upper bound $U\subt{max}$ for the potential $U_E$ in this region. For three degenerate neutrinos it is given by eq.~\eqref{AQ6}, with calculable modifications for arbitrary masses of the neutrinos. It also predicts for the standard model coupled to gravity that $U_E$ remains negative for the whole range of $\rhotil$ below a value close to the maximum of $U_E$. The minimum of $U_E$ occurs in this case at $\rhotil=0$ for negative $U_E$. It is difficult to see how realistic cosmology emerges in this case within the truncation~\eqref{C3}. In contrast, for grand unified theories $\rhotil=0$ corresponds to the maximum of $U_E$, predicting an inflationary epoch. The restrictions for the UV-fixed points of $u(0)$ and $f(0)$ remain valid for arbitrary renormalizable theories since these are also the UV-fixed point values for the flow with $k$. For $k\to0$ these restrictions are typically no longer present for general renormalizable theories, however. They can be circumvented, at least partially, by the flow away from the scaling solution.

\subsection{Solution of cosmological constant problem}\label{subsec:SCCP}

A central reason why we focus this work on the scaling solution is the possible dynamical solution of the cosmological constant problem~\cite{WEICC} without invoking any small dimensionless parameter. This distinguishes the scaling solution or close neighbors of it from the more general solutions of flow equations for quantum gravity.

The central ingredient for this dynamical solution is the decrease of $\Vhat(\rhotil)$ to zero for $\rhotil\to\infty$. If the cosmological dynamics drives $\rhotil$ towards infinity in the infinite future, the potential in the Einstein frame $U_E=\Vhat M^4$ vanishes for $t\to\infty$. The cosmological constant is driven to zero dynamically. More precisely, only the dimensionless ratio of the scalar potential $U$ over the fourth power of the Planck mass $F^2$ is observable. This ratio vanishes due to $U(\rho\to\infty)\sim k^4$, $F(\rho\to\infty)\sim\rho$, such that $\Vhat=U/F^4\sim\rho^{-2}$. For the scaling solution this is directly visible in the relation
\bel{Q1A}
\Vhat(\rhotil\to\infty)=\frac{u}{f^2}(\rhotil\to\infty)=\frac{u_\infty^2}{4\xi_\infty^2\rhotil^2}\ .
\ee

For the scaling solution the only mass scale is given by the renormalization scale $k$. The large present value of $\rhotil$ is only due to the evolution of the universe - in Planck units the present universe is very old. This large value explains the tiny value of the present dark energy density in Planck units, $\Vhat(\text{today})\approx10^{-120}$. No small parameter or small ratio of parameters is invoked, similar to the present small value of the matter and radiation energy density in Planck units, which also results from the large age of the universe.

\subsection{Relevant parameters}

In order to appreciate the role of fundamental scale invariance for the dynamical solution of the cosmological constant problem we compare with general renormalizable quantum gravity. Indeed, the situation is different for general solutions of the flow equation with relevant parameters. Typically both the Planck mass and the cosmological constant correspond to relevant parameters. The general solution of the flow equation takes then the qualitative form
\begin{align}
\label{CC1}
F=&f(\rhotil)k^2+\mu_p^2\approx f_0k^2+\xi_\infty\chi^2+\mu_p^2\ ,\nn\\
U=&u(\rhotil)k^4+\lambda\ ,
\end{align}
where both $\mu_p^2$ and $\lambda$ set intrinsic mass scales related to relevant parameters. These are free parameters of the model. (More generally, $\mu_p^2$ and $\lambda$ may be functions of $\chi$ which only depend on two constants that we associate here to the constants $\mu_p^2$ and $\lambda$. The computation of these functions depends on the flow in a range of $k$ were deviations from the scaling solution grow large.)

If the present value of $F$ is dominated by $\mu_p^2$, i.e. $\xi\chi^2\lesssim\mu_p^2$, one finds $\Vhat\approx\lambda/\mu_p^4$. A small value of $\Vhat$ corresponds now to a ratio of parameters that has to assume the tiny value $10^{-120}$. This requires \qq{tuning} of the relevant parameters. This situation is also realized in the absence of a scalar field $\chi$, or if $\chi$ settles in the cosmological evolution to a constant value $\bar\chi=c\mu_p$. In the latter case one replaces in eq.~\eqref{CC1} $\mu_p^2$ by $\mu_p^2(1+\xi c^2)$.

Without a tuning of relevant parameters i.e. for $\lambda\approx\mu_p^4$, a dynamical solution of the cosmological constant problem remains still possible for crossover cosmologies leading to a present value of $\chi\approx10^{30}\mu_p$. As compared to the scaling solution one replaces $f_0k^2\to f_0k^2+\mu_p^2$, $uk^4\to uk^4+\lambda$. This is the case discussed in sect.~\ref{sec:QG} for which the scaling solution becomes relevant only for $k^2\gtrsim\mu_p^2$. Indeed, in this range of $k$ we can neglect $\mu_p^2$ and $\lambda$ in eq.~\eqref{CC1}. While the relevant parameters $\mu_p^2$ and $\lambda$ become unimportant in the UV-limit $k\to\infty$, they still influence the early cosmology for small values of the scalar field $\chi$. The reason is that the limits $k\to\infty$ and $\chi\to0$ can no longer be identified as for the exact scaling solution.

In the presence of relevant parameters the effective scalar potential in the Einstein frame becomes
\bel{189}
U_E=\frac{\gl u(\rhotil)k^4+\lambda\gr M^4}{\gl f_0k^2+2\xi_\infty\rhotil+\mu_p^2\gr^2}\ ,
\ee
with $u(\rhotil)$ the scaling potential. This potential can be positive for $\rhotil=0$ even for $u_0<0$, provided $\lambda>-u_0k^4$. In the presence of relevant parameters an inflationary epoch may become possible for the standard model of particle physics coupled to gravity. On the way towards $k\to0$ our approximation at $\rhotil=0$ break down since $v<1$ requires $2\gl\lambda+u_0k^4\gr<\mu_p^2k^2+f_0k^4$. For an inflationary epoch the scale $k$ may, however, be replaced by an effective geometrical cutoff.

If $u(\rhotil)+\lambda/k^4$ is positive for the whole range of $\rhotil$, a crossover cosmology with $\chi$ diverging in the infinite future is rather generic. The potential $U_E(\vp)=M^4u/f^2$ typically decreases monotonically for increasing $\vp$ due to the exponential increase of $f$. For such a potential $\vp$ increases monotonically with time, at least asymptotically. In the infinite future one reaches $U_E(\vp\to\infty)=0$. This situation is realized for the scaling solution of pure gravity coupled to the scalar field $\chi$. For particle physics with the standard model as an effective low energy theory the situation is more complex. For fundamental scale invariance the potential $U_E$ is now negative for a substantial range of $\chi$. The asymptotic cosmological scaling solution~\eqref{DS6} is now only reached if the increase of $\chi$ during the kination epoch is large enough. This poses conditions on $Z(\vp)$. This issue can be avoided for the more general solution of the flow equations~\eqref{CC1} provided that $\lambda+k^4u(\rhotil)$ is positive for the whole range of $\rhotil$. The price to pay is a partial loss of predictive power.

We conclude that the basic ingredients for a dynamical solution of the cosmological constant problem are similar for general renormalizable quantum gravity without fine tuning of parameters (i.e. for $\lambda$ of the order $\mu_p^4$) and for fundamental scale invariance. This requires, however, that the scale $\mu_p$ set by the relevant parameters is much smaller than the present value of the Planck mass given by $\chi$, typically $\mu_p\approx10^{-2}\,\text{eV}$. In both cases late cosmology corresponds to large $\chi$ such that $F$ is dominated by $\xi_\infty\chi^2$. On the other hand, $U$ becomes a constant $U_\infty$ for large $\chi$, resulting in $U_E\sim U_\infty M^4/\gl\xi_\infty^2\chi^4\gr$. As a result, $U_E$ vanishes as $\chi$ increases towards infinity. For cosmology of late stages of inflation and all later epochs only the relevant parameter $\lambda$ distinguishes general renormalizable gravity from the setting of fundamental scale invariance.

\section{Discussion}\label{sec:D}

A formulation of quantum gravity as a quantum field theory for the metric, together with the power of modern functional renormalization group techniques to compute the effects of quantum fluctuations of the metric, yields a predictive scheme for cosmology. Important functions as the inflaton potential for early cosmology, or the cosmon (quintessence) potential for late cosmology, can no longer be chosen completely freely. The scaling solution which is necessary to render the quantum field theory renormalizable imposes important restrictions on the shape of these potentials. At low energies and for small field values renormalizability enforces for the Higgs scalar and other scalars with gauge or Yukawa interactions to the fields of the standard model an approximately polynomial potential. This is not the case for the scalar singlet discussed in this work. The scaling solution predicts a potential that deviates strongly from a polynomial form. The potential in the Einstein frame, or the frame invariant potential, approaches zero exponentially for large positive values of the canonical scalar field $\sigma$, while it tends to a positive constant in the opposite limit of large negative $\sigma$.

It is impressive to see how from these simple properties an overall cosmology with rather realistic features emerges. Cosmology describes a crossover from a UV-fixed point, realized for $\sigma\to-\infty$ in the infinite past, to an IR-fixed point for $\sigma\to\infty$ in the infinite future. In-between, the sequence of characteristic cosmological epochs finds its place: inflation, kination, radiation domination, matter domination and dark energy domination.

The rather detailed information about the scaling solution yields further predictions for the case of fundamental scale invariance.
\begin{enumerate}[(i)]
\item Within the truncation with up to two derivatives the standard model coupled to quantum gravity is not viable. This is related to the negative value of the scaling potential at $\rhotil=0$, $u_0<0$. The conclusion remains the same if we replace the scalar singlet by the Higgs doublet, as for Higgs inflation. A viable quantum gravity extension of the standard model requires an important role of higher derivative terms~\cite{GORS}. We pursue here the alternative of an extension to a grand unified symmetry. This implies positive $u_0$ for which an inflationary epoch is predicted. 
\item Inflation ends at the latest at the transition where many bosons beyond the standard model particles become massive. This is typically the GUT-phase transition in grand unified theories. An end of inflation near the GUT-phase transition can naturally explain the small amplitude of the primordial fluctuations.
\item Dark energy is dynamical rather than being a cosmological constant.
\item For the range of the scalar field relevant after inflation the potential is bounded by a maximal value $U\subt{max}=(0.32m\nb)^4$, or similar for a non-degenerate mass pattern of the three neutrinos. The neutrino mass $m\nb$ sets the characteristic scale for the dynamics of dark energy. The maximal potential $U\subt{max}$ limits the amount of a possible early dark energy~\cite{EDE1, EDE2, GZAV, DLSW, DOS, PSKK, NSS, YEPI, GVZH} unless additional degrees of freedom are introduced.
\end{enumerate}

So far we have not used quantitative information about the wave function renormalization or kinetial $Z(\vp)$ or $K(\chi)$. Once the scaling solution is computed for this function important additional predictions become possible. For example, there is an explicit formula for the slow roll coefficients $\eps$ and $\eta$ during inflation in terms of the three scaling functions. Since these coefficients determine directly the spectrum of primordial cosmic fluctuations such a prediction is testable. It is remarkable that cosmological observations will be able to falsify a model of fundamental scale invariance with a single scalar field. Future computational progress for the functional flow equations and the associated scaling functions may lead to a distinction which type of models for momenta near the Planck scale a viable or not.

For more general renormalizable quantum field theories of gravity the predictive power is somewhat reduced due to the presence of relevant parameters that can be chosen freely. Still, only a small number of such parameters characterizes the possible deviations from the scaling solution. Also for this case we expect substantial additional constraints for cosmology once the scaling form of the kinetial is computed.

\nocite{*}
\bibliography{refs}
\end{document}